\documentclass[12pt]{article}
\usepackage{amsmath} 
\usepackage{multirow} 
\usepackage{amsfonts}
\usepackage{amssymb,graphics,psfrag}
\usepackage{array,epsfig,multirow,stmaryrd,graphicx,mathtools}
\usepackage{comment}
\usepackage{hyperref}
\def\hybrid{\topmargin -20pt    \oddsidemargin 0pt
        \headheight 0pt \headsep 0pt
        \textwidth 6.25in       
        \textheight 9 in       
        \marginparwidth .875in
        \parskip 5pt plus 1pt 
          \jot = 1.5ex
   }
\hybrid
\numberwithin{equation}{section}
\numberwithin{table}{section}

\newcommand{\beq}{\begin{equation}}
\newcommand{\eeq}{\end{equation}}
\newcommand{\be}{\begin{equation}}
\newcommand{\ee}{\end{equation}}
\newcommand{\bea}{\begin{eqnarray}}
\newcommand{\eea}{\end{eqnarray}}   
\newcommand{\ben}{\begin{eqnarray*}}
\newcommand{\een}{\end{eqnarray*}}                  
\newcommand{\ba}{\begin{aligned}}
\newcommand{\ea}{\end{aligned}}
\newcommand{\bt}{\begin{tabular}}
\newcommand{\et}{\end{tabular}}
\newcommand{\bc}{\begin{center}}
\newcommand{\ec}{\end{center}}

\newcommand{\cO}{\mathcal{O}}
\newcommand{\cT}{\mathcal{T}}

\newcommand{\cC}{\mathcal{C}}
\newcommand{\cD}{\mathcal{D}}
\newcommand{\cL}{\mathcal{L}}

\newcommand{\cK}{\mathcal{K}}
\newcommand{\cN}{\mathcal{N}}

\newcommand{\cG}{\mathcal{G}}
\newcommand{\cA}{\mathcal{A}}

\newcommand{\cB}{\mathcal{B}}
\newcommand{\cF}{\mathcal{F}}

\newcommand{\cR}{\mathcal{R}}

\newcommand{\cV}{\mathcal{V}}

\newcommand{\I}{\text{Im}}
\newcommand{\R}{\text{Re}}

\newcommand{\bbZ}{\mathbb{Z}}
\newcommand{\bbR}{\mathbb{R}}
\newcommand{\bbC}{\mathbb{C}}
\newcommand{\bbP}{\mathbb{P}}

\newcommand{\nn}{\nonumber}

\newcommand{\RE}{\textrm{Re} \,}

\newcommand{\cref}{{\bf [check ref]}}

\psfrag{n1}{$\nu_1$}
\psfrag{n2}{$\nu_2$}
\psfrag{n1'}{$\nu_1'$}
\psfrag{n2'}{$\nu_2'$}
\psfrag{n9}{$\nu_9$}
\psfrag{n10'}{$\nu_{10}'$}
\psfrag{t1}{$\tilde{\nu}_1$}
\psfrag{t2}{$\tilde{\nu}_2$}
\psfrag{t9}{$\tilde{\nu}_9$}
\psfrag{t1'}{$\tilde{\nu}_1'$}
\psfrag{t2'}{$\tilde{\nu}_2'$}
\psfrag{t10'}{$\tilde{\nu}_{10}'$}

\def\blfootnote{\xdef\@thefnmark{}\@footnotetext} 
\long\def\symbolfootnote[#1]#2{\begingroup%
\def\thefootnote{\fnsymbol{footnote}}\footnote[#1]{#2}\endgroup}

\begin{document}

\baselineskip=17pt

\begin{titlepage}
\begin{flushright}
\parbox[t]{1.1in}{
MPP-2011-107}
\end{flushright}

\begin{center}

\vspace*{ 1.2cm}

{\large \bf Gravitational Instantons and Fluxes from M/F-theory  \\[.1cm]  on Calabi-Yau fourfolds
}

\vskip 1.2cm

\renewcommand{\thefootnote}{}
\begin{center}
 {Thomas W. Grimm and\ \,Raffaele Savelli  \footnote{grimm,\ savelli\ \textsf{at}\ mpp.mpg.de}}
\end{center}
\vskip .2cm
\renewcommand{\thefootnote}{\arabic{footnote}} 

{Max-Planck-Institut f\"ur Physik, \\
F\"ohringer Ring 6, 80805 Munich, Germany} 

 \vspace*{1.5cm}

\end{center}

\vskip 0.2cm
 
\begin{center} {\bf ABSTRACT } \end{center}

We compactify four-dimensional $\cN=1$ gauged supergravity theories 
on a circle including fluxes for 
shift-symmetric scalars. Four-dimensional Taub-NUT gravitational instantons 
universally correct the three-dimensional superpotential in the absence of fluxes.
In the presence of fluxes these Taub-NUT instanton contributions
are no longer gauge-invariant.  Invariance can be restored by  
gauge instantons on top of Taub-NUT instantons. We establish the embedding 
of this scenario into M-theory. Circle fluxes and gaugings arise
from a restricted class of M-theory four-form fluxes on a resolved  Calabi-Yau fourfold. 
The M5-brane on the base of the elliptic fourfold dualizes into the universal 
Taub-NUT instanton. In the presence of fluxes this M5-brane is anomalous.
We argue that anomaly free contributions arise from involved M5-brane geometries dual
to gauge-instantons on top of Taub-NUT instantons.
Adding a four-dimensional superpotential to the 
gravitational instanton corrections leads to  three-dimensional Anti-de Sitter vacua at 
stabilized compactification radius. 
We comment on the possibility to uplift these M-theory vacua, 
and to  tunnel  to four-dimensional F-theory vacua.

\end{titlepage}

\tableofcontents

\vspace*{2cm}

%
%

\section{Introduction}

The study of four-dimensional supergravity theories within string theory has been approached in 
various compactification schemes. A rather general goal of these approaches is to understand
the set of supergravity theories arising in a consistent quantum theory of gravity.  
Ideally one would thus be able to identify either a restrictive class of consistency 
conditions, or, even more ambitious, discover a dynamics which prefers certain string reductions.
A more modest aim is to identify the concepts and tools required to study such four-dimensional (4d)
effective theories. In this work we investigate 4d, $\cN=1$ supergravity
theories from the perspective of a quantum corrected three-dimensional (3d) effective 
theory obtained by a flux compactification on a circle. 

We start our investigation with a class of 4d, $\cN=1$ supergravity actions describing
a set of complex scalars together with a non-Abelian gauge theory.  Of particular interest
will be cases where the imaginary parts of the scalars posses shift symmetries. 
We will include terms quadratic in the curvature in our investigation.
Such effective 
theories naturally arise by compactifying F-theory on singular Calabi-Yau fourfolds. 
The singularities correspond to space-time filling 7-branes supporting 
the non-Abelian gauge groups. Switching on two-form fluxes on the world volume
of the 7-branes can induce a minimal gauging of part of the shift symmetries. 
This generically breaks the gauge group to the Abelian Cartan torus. 
In such theories we study two types of 4d background configurations 
in detail: (1) a 3d Minkowski space times a circle with    
non-trivial field-strength of the shift-symmetric scalars, 
(2) a 4d gravitational instanton solution which might admit a 
4d gauge instanton bundle. Let us comment on these two configurations in turn.

A circle compactification with non-trivial vacuum expectation value 
for the field strength of the shift-symmetric 
scalars can be viewed as the simplest flux compactification \cite{Blumenhagen:2006ci,Douglas:2006es}. 
Using a Kaluza-Klein reduction we determine the 3d gauged supergravity
theory and show that it contains gaugings inherited from 4d, as well as, new 
genuinely 3d gaugings induced by the fluxes around the circle. In 
three dimensions all degrees of freedom can be encoded by dynamical scalars 
and non-dynamical vectors encoding the gauging. The 3d, $\cN=2$ supergravity theory 
formulated in this language will be our starting point to study the scalar potential, determine 
flux vacua, and discuss quantum corrections. These effective phenomena are then 
argued to have an M-theory interpretation which is inferred by the duality between 
M- and F-theory \cite{Vafa:1996xn,Denef:2008wq,Weigand:2010wm}.  
We link the above introduced 3d supergravities 
to an M-theory reduction on an elliptically fibered Calabi-Yau fourfold with 
a non-trivial flux for the M-theory four-form field strength. In our compactification schemes we will always neglect the effects of a non-trivial warp factor.

The gravitational instantons under consideration are solutions to the Euclidean Einstein 
equations with anti-self-dual two-form curvature tensor. 
More precisely, we will focus on a special class of solutions,
known as Taub-NUT spaces \cite{Eguchi:1980jx}.
The Taub-NUT spaces have finite action and are believed 
to contribute new saddle-points of the 4d path integral summing over gravity backgrounds
with different topologies \cite{Gibbons:1994cg}. 
However, 
instead of computing directly in the four dimensions we will include the 
Taub-NUT instanton corrections in the effective three-dimensional supergravity theory
obtained by the fluxed circle reduction. While Taub-NUT instantons 
generically break 4d supersymmetry in the background, they can supersymmetrically
correct the 3d effective supergravity theory.
This has a natural analog in the M-theory picture where the Taub-NUT instantons are 
identified with M5-brane instantons on the base of the elliptic fibration. In fact, such 
M5-branes naturally contribute to the 3d effective superpotential in the absence of fluxes~\cite{Witten:1996bn}.

To be more precise, we will find that in the absence of 3d gaugings the Taub-NUT gravitational 
instantons yield a universal contribution to the 3d superpotential proportional to $e^{-2 \pi T_0}$. Here the complex scalar 
$T_0$ parameterizes  the square of the radion, i.e.~the radius of 
the circle, complexified with the scalar obtained by dualizing the Kaluza-Klein vector in 
the 4d metric. This induces a potential for the radion which can lead to new supersymmetric 
AdS$_3$ vacua if already  a 4d superpotential $W_0$ is present. The supersymmetric stabilization 
in an AdS vacuum is the lower-dimensional analog to the proposal of 
KKLT to stabilize K\"ahler moduli~\cite{Kachru:2003aw}. In this new vacuum the theory can be 
effectively three-dimensional and stable. A natural conjecture is to up-lift this setup 
to positive vacuum energies. This generates a meta-stable 3d de Sitter vacuum which can tunnel 
to a 4d de Sitter vacuum. Such transdimensional tunnelings have been discussed, for example, in  \cite{Linde:1988yp,Giddings:2004vr,BlancoPillado:2009di,Carroll:2009dn,Graham:2010hh}.
In this work we will rather focus on the first steps in implementing such a scenario. It turns 
out that the combination of the fluxed reduction and Taub-NUT instanton correction already 
imposes interesting compatibility conditions.

If we consider the complete flux compactification of the 4d gauged supergravity theory, the corrections
due to Taub-NUT gravitational instantons will be altered. We show that the fluxes along the 
compactification circle enforce that the shift symmetry of the imaginary part of 
the radion modulus $T_0$ is gauged. This forbids a superpotential $\cA e^{-2 \pi T_0}$, when the 
pre-factor $\cA$ does not also transform under the $U(1)$ gauge symmetry. However, we find that one can 
indeed include a field-dependent $\cA$ to cancel the shifts by using the gaugings already present 
in the 4d theory, together with fluxes on the Taub-NUT instanton. This yields a superpotential of
the form recently discussed in \cite{Grimm:2011dj}, in which two scalars with gauged shift symmetries 
are appropriately combined using fluxes on the instantons. 
In the microscopic M-theory picture the absence of a superpotential with constant $\cA$ has
an anomaly interpretation. In the presence of fluxes the M5-brane on the base of the elliptic fibration becomes 
anomalous \cite{Witten:1999vg}, in a way which is familiar from the  
Freed-Witten anomaly for D-branes \cite{Freed:1999vc}. Only for a more complicated 
M5-brane worldvolume, on which the 
four-form fluxes encoding the 4d gaugings vanish, this anomaly can be absent. 
On the one hand, this is precisely the string 
dual to a Type IIB compactification with a Taub-NUT instanton supporting  
gauge bundles. This is a realization of our original 4d configuration. On the other hand, 
in the Type IIA limit of M-theory this setup can be interpreted in terms of D4-D6-NS5 branes, 
as we discuss briefly.

The paper is organized as follows:
In section~\ref{from4to3} we perform the dimensional reduction from 4d to 3d on 
a circle by allowing for a flux background for the shift-symmetric scalars, and discuss the inclusion of the Taub-NUT instantons
from a 3d perspective. In subsection~\ref{3dlandscape} the Kaluza-Klein reduction is discussed 
in the absence of 4d vectors. We also introduce the 3d $\cN=2$ supergravities with dynamical scalars and 
non-dynamical 3d gauge vectors. In subsection~\ref{4dVectors} we generalize this reduction by introducing 4d vectors and gaugings
of 4d shift symmetries. We determine the combined scalar potential. The stabilization of the radion modulus by Taub-NUT gravitational instantons
is discussed in subsection~\ref{StabRadion}. We motivate the form of the Taub-NUT corrections by studying the instanton action. Furthermore, we comment
on a KKLT-like scenario to stabilize the size of the compactification circle, and briefly speculate on the up-lift and tunneling from 3d to 4d.
The form of the superpotential in the presence of circle fluxes, vectors and 4d gaugings is discussed in subsection~\ref{FluxSuperpotential}. 
This requires the introduction of gauge instantons on top of the gravitational instanton.

In section~\ref{stringinterpret} we provide the M-theory construction 
of the 3d effective theory and combine fluxes and instanton corrections. 
We discuss M-theory on elliptically fibered Calabi-Yau fourfolds in 
subsection~\ref{mod_id}. We also introduce its F-theory limit in which the M-theory elliptic fiber 
shrinks to zero size which yields in the T-dual picture to a growing  
large extra dimension. The resulting setup can be matched with an 
F-theory compactification. Four-form fluxes in M-theory are introduced 
in subsection~\ref{Mfluxes}, where we 
also comment on their four-dimensional interpretation.  
The resulting 3d D-term potential is introduced in subsection~\ref{Minterpret}. 
The identification of the Taub-NUT gravitational instanton as M5-brane is discussed 
in subsection \ref{Taub-NUT_M}. We also study the interplay 
of four-form fluxes and the M5-brane instantons.


\section{Circle flux reductions and gravitational instantons}\label{from4to3}

In this section we analyze the compactification of 4d, $\cN=1$ supergravity 
theories to three dimensions on a circle 
by including fluxes for the field strengths of shift-symmetric scalars in subsections~\ref{3dlandscape} and \ref{4dVectors}.
The gravitational instantons, and their contribution to the 3d superpotential are discussed in subsections~\ref{StabRadion} and \ref{FluxSuperpotential}.
Section~\ref{from4to3} will be entirely formulated in the language of the supergravity theories, and hence can 
be understood independently of the M-theory realization discussed 
in section~\ref{stringinterpret}. However, the structure of certain 4d couplings will be familiar from a F-theory compactifications
to four space-time dimensions \cite{Grimm:2010ks,Grimm:2011tb,GrimmTaylor} which will later make the embedding into M-theory more immediate.

\subsection{Fluxed reduction from 4d to 3d}\label{3dlandscape}

In this subsection we will discuss how 
non-trivial fluxes can arise in the reduction on a circle and generate a landscape of vacua in the effective three-dimensional theory. 
In our discussion we will generalize the following simple observation. Consider a real scalar $\phi$ which only appears with 
a derivative $F_1 = d\phi$ in the four-dimensional action. In the $S^1$ reduction from four to three dimensions
$F_1$ can acquire a background flux $M = \int_{S^1}\langle F_1\rangle$. The constants $M$ will induce a scalar potential 
in the three-dimensional effective theory. Note that this is the simple analog of the more complicated flux compactifications from 
ten to four dimensions reviewed in \cite{Blumenhagen:2006ci,Douglas:2006es}.

Let us include such fluxes when dimensionally reducing four-dimensional $\cN=1$
theories. 
In this section we will assume 
that there are no 4d vectors in the spectrum, and postpone a more general discussion to the next subsection.
The purely bosonic part of the 4d supergravity action reads
\beq\label{4dkineticphi}
S^{(4)}=- \int \tfrac12\,R_4\, *_4{\bf{1}}+ K^{\rm 4d}_{T_\alpha \bar T_{\bar\beta}} d T_\alpha \wedge *_4{d \bar  T_\beta}  + V_{\rm F} *_4{\bf 1} + \cL_{\cR^2} \ ,
\eeq
with 
\beq \label{def-cLR2}
  \cL_{\cR^2} = \tfrac12  {\rm Re} \, \sigma \, {\rm Tr}(\cR_4 \wedge *_4 \cR_4)
+\tfrac12 {\rm Im}\, \sigma \, {\rm Tr} ( \cR_4 \wedge \cR_4 )\ ,
\eeq
where $*_4$ is the 4d Hodge star, and $\cR_4 = \frac12  R_{\mu \nu} dx^\mu \wedge dx^\nu$ is the $SO(3,1)$-valued curvature two-form. Note that we have included the higher curvature terms which 
are quadratic in the curvature. 
The inclusion of such corrections has been argued to be crucial in linking 
the effective theory and the stingy compactification geometries \cite{GrimmTaylor}. 

In order to perform the reduction we will assume that some of the real four-dimensional scalar fields $ \I T_\alpha$ 
have global shift symmetries. These will later be gauged in section~\ref{4dVectors}.
We denote the number of independent shift symmetries by $n_s$.
This implies that $n_s$ combinations of the fields $ \I T_\alpha$ only appear with derivatives in the 4d action.
To specify the fields $\I T_\alpha$ which have a shift symmetry 
we introduce the constant `symmetry tensor' $ Q_{\alpha}^\Sigma$ such that
\beq \label{shift-symmetry}
    \I T_\alpha \ \rightarrow \ \I T_\alpha + Q_\alpha^\Sigma\, \Lambda_\Sigma\ , \qquad \Sigma= 1\ldots n_{s}
\eeq
for constants $\Lambda_\Sigma$, is a symmetry of the effective action \eqref{4dkineticphi}. To specify the 
shift symmetries we introduced a general $Q_{\alpha}^\Sigma$ with fixed rank $n_s$. Clearly, upon choosing
an appropriate basis of the $T_\alpha$ one can set  $Q_{\alpha}^\Sigma= (\delta^\Sigma_\Lambda,0)$ such that the first $n_s$ $T_\alpha$'s have 
shift symmetries.
There are several reasons for a non-maximal rank of $Q_{\alpha}^\Sigma$. 
The K\"ahler metric $K^{\rm 4d}_{T_\alpha \bar T_{\bar\beta}}$, which integrates to a  
 K\"ahler potential $ K^{\rm 4d}$, can explicitly depend on $ \I T_\alpha$ breaking the
shift symmetry. However, note that $K^{\rm 4d}$ is non-holomorphic and hence can consistently 
depend on $\R T_\alpha$ only. The latter fact should be contrasted with the holomorphic 
 superpotential $W$ which 
specifies the scalar potential $V_{\rm F}$ in 
the action \eqref{4dkineticphi} via
\beq
     V_{\rm F} = e^{K^{\rm 4d}} \big(K_{\rm 4d}^{T_\alpha \bar T_\beta}\,  D_{T_\alpha} W_{\rm 4d} \, \overline{D_{T_\beta} W_{\rm 4d}} -3 |W_{\rm 4d}|^2\big)\ .
\eeq
The superpotential breaks the shift symmetry as soon as it depends on the corresponding $T_\alpha$.
A well-known example for a non-trivial $W_{\rm 4d}$ are superpotentials of the schematic form $W_{\rm 4d}= W_0 + A\, e^{-n^\alpha T_\alpha}$.
The $T_\alpha$-dependent term arises in string theory compactifications from non-perturbative effects, and clearly breaks the 
shift-symmetry of the corresponding $n^\alpha \I T_\alpha$. Such non-perturbative contributions are essential in 
the study of moduli stabilization in the 4d effective theory as argued in \cite{Kachru:2003aw,Blumenhagen:2006ci,Douglas:2006es}.

In addition to $K^{\rm 4d}$ and $W_{\rm 4d}$ also a field 
dependent coupling $\sigma(T)$ in front of the higher curvature terms can in principle break the shift symmetry. 
In string and M-theory compactifications, as we discuss below, a simple form of $\sigma$ for large $\R T_\alpha$ is \cite{GrimmTaylor}
\beq  \label{simple_sigma}
  \sigma = \tfrac18 k^\alpha T_\alpha  \ ,
\eeq
for a constant vector $k^\alpha$. Since $Q^\Sigma_\alpha$ parameterizes the shift invariant $\I T_\alpha$ appearing in the 
action, a coupling \eqref{simple_sigma} in the effective action suggests the natural restriction  of $Q^\Sigma_{\alpha}$: 
\beq \label{ktheta}
  k^\alpha Q_\alpha^\Sigma = 0\ .
\eeq 
Strictly speaking, for a gravitational coupling with \eqref{simple_sigma} the action can be made invariant 
under the shift symmetry of $\I \sigma$ if boundary terms are properly taken into account, and one uses the gravitational 
Chern-Simons form. However, it turns out that assuming the absence of a shift 
symmetry for $\I \sigma$ simplifies the analysis significantly, and we will make the assumption \eqref{ktheta} in the following.
It will be interesting to see in the
setups of sections \ref{stringinterpret} how this condition is realized in M-theory.

We now want to reduce such a theory to an three-dimensional $\mathcal{N}=2$ supergravity theory 
on a background geometry of topology $\mathbb{R}^{1,2}\times S^1$ with metric
\be\label{backmetric}
\langle g^{(4)}\rangle=\left(\begin{array}{cc}\eta^{(3)}_{\mu\nu}&0\\0&\langle r^2\rangle\end{array}\right)\,,
\ee
$\langle r\rangle$ being the vev of the radion field which parameterizes the circumference of $S^1$. In this background 
we also allow for a non-trivial profile of the ${\rm Im}\,T_\alpha$ along the $S^1$ which posses a 
shift symmetry \eqref{shift-symmetry}. This allows us to include $n_s$ 
independent background fluxes $M_\Sigma$ as
\be\label{fluxImT}
M_\alpha\equiv Q^\Sigma_\alpha M_\Sigma = \int_{S^1}d\langle \I T_\alpha\rangle\,.
\ee
Note that this form of $M_\alpha$ implies together with \eqref{ktheta} that
\be\label{kthetaM}
k^\alpha M_\alpha =0\,.
\ee
Clearly such a flux should be proportional to the unique harmonic one-form on $S^1$, which with a slight abuse of notation may 
be written in the form $dy$.\footnote{Note that one would have to introduce 
two patches on $S^1$ to give a proper global definition of the harmonic one-form.} Hence \eqref{fluxImT}  implies that the local expression 
for $\langle{\rm Im}T_\alpha\rangle$ is, up to an additive constant, $\langle{\rm Im}T_\alpha\rangle=M_\alpha\, y$. 
So far we did not give any reason for the $M_\alpha$ or $M_\Sigma$ to obey a quantization rule, and thus, for now, they are real numbers. 
However, we will see in subsection \ref{Minterpret} that their M-theory origin forces them to satisfy some quantization condition dictated by global consistency constraints.

The three-dimensional effective theory is computed by including the fluctuations around this background.
The fluctuation of the metric  takes the following generic form
\be\label{metricfluct}
g^{(4)}(x^{\mu})=\left(\begin{array}{cc}g^{(3)}_{\mu\nu}+r^2A^0_\mu A^0_\nu &r^2A^0_\mu\\ r^2A^0_\nu&r^2\end{array}\right)(x^{\mu})\,,
\ee
where $g^{(3)}$ is the fluctuation of the 3d metric in the Einstein frame and $A^0$ is the Kaluza-Klein vector of the reduction.
Restricting to the lowest modes $r,A^0$ and $g^{(3)}$ are only functions of the 3d coordinates $x^\mu$. Note that for the reduction defined by \eqref{metricfluct}, 
the real scalar accompanying $A^0$ in a 3d, $\cN=2$ vector multiplet will be $r^{-2}$, and we define 
\beq
   R = r^{-2}\ .
\eeq
We also include the fluctuations of the four-dimensional fields $T_\alpha$ which correspond to three-dimensional complex scalars. By abuse 
of notation we also denote them by $T_\alpha$ keeping in mind that they only depend on the 3d coordinates $x^\mu$
when used in a three-dimensional context.

Performing the reduction, the kinetic term in \eqref{4dkineticphi} simply reduces 
to the usual 3d kinetic term for the fields $T_\alpha$ promoted to 3d scalar fields plus a scalar potential.
After integration over $S^1$ and performing the Weyl rescaling $g^{(3)}\rightarrow r^{-2}g^{(3)}=Rg^{(3)}$ in order to bring the 3d Einstein-Hilbert term to the canonical form, 
we find
\bea\label{3dkineticT}
S^{(3)}&=& - \int \tfrac12 R_3*_3{\bf{1}}-\tfrac14\tilde K_{RR} \big(F^0\wedge *_3 F^0 + dR\wedge *_3 dR\big) +  \tilde K_{T_\alpha \bar T_{\bar\beta}} \nabla T_\alpha\wedge *_3  \overline{\nabla T_{\beta}} \nn \\ 
&&\qquad + \big( R^2\,\tilde K_{T_\alpha \bar T_{\bar\beta}} M_\alpha M_\beta + R^2 V_{\rm F}\big)\, *_3\bf{1} + \tilde \cL_{\cR^2} \ ,
\eea
where now $*_3, R_3$ are respectively the Hodge star and Ricci scalar with respect to the 3d Einstein frame metric. $\tilde \cL_{\cR^2}$ is a complicated expression 
for the reduction of $\cL_{\cR^2}$ given in \eqref{def-cLR2}, and we will not need this form in the following.    
The negative definite metric $\tilde K_{RR}$ and the positive K\"ahler metric $\tilde K_{T_\alpha \bar T_{\bar\beta}}$ 
can be determined from the function $\tilde K$, the kinetic potential, of the form
\be\label{KineticPotential}
\tilde K(T_\alpha+\bar T_\alpha|R)=\log R+K^{\rm 4d}(T_\alpha +\bar T_\alpha)+1\,.
\ee
Note that 
due to the non-trivial background fluxes \eqref{fluxImT} the ordinary derivatives in \eqref{4dkineticphi} are reduced 
to  the three-dimensional covariant (or rather invariant) derivatives 
\beq\label{InvDerivat}
     \nabla T_\alpha = d T_\alpha + iM_\alpha A^0\ .
\eeq
This implies that even in the absence of a gauging in four dimensions, the three-dimensional theory will be a gauged supergravity theory.  In this derivation we have used the inverse of the metric fluctuation \eqref{metricfluct}, which looks like
\be\label{invmetricfluct}
g^{(4)\,-1}(x^{\mu})=\left(\begin{array}{cc}g^{(3)\,\mu\nu} &-g^{(3)\,\mu\nu}A^0_\nu\\ -A^0_\mu g^{(3)\,\mu\nu}&\frac{1}{r^2}+g^{(3)\,\mu\nu}A^0_\mu A^0_\nu\end{array}\right)(x^{\mu})\,.
\ee 
It is easy to see that the scalar potential in \eqref{3dkineticT} arises from the $1/r^2$ part of the last component of \eqref{invmetricfluct}. The $A^0$-dependent part of \eqref{invmetricfluct}, instead, is responsible for the appearance of the covariant derivative acting on the $T_\alpha$.

The gaugings generated by the flux $M_\alpha$ have a direct geometrical interpretation in terms of 
the  1-dimensional diffeomorphisms of $S^1$, which, after reduction, turns into a gauge symmetry of 
the 3d theory with gauge boson $A^0$. Indeed, by the Kaluza Klein Ansatz \eqref{metricfluct}, the 
coordinate transformation $y\rightarrow y+\lambda^0(x^\mu)$ induces the gauge transformation 
$A^0_\mu\rightarrow A^0_\mu-\partial_\mu\lambda^0$. Now, since we required an explicit dependence on 
$y$ of the 4-dimensional field ${\rm Im}T_\alpha$, the corresponding 3d fluctuations ${\rm Im}T_\alpha(x^\mu)$ 
acquire a charge with respect to such a symmetry. In fact, one has for the 4d scalars:
\be
{\rm Im}T_\alpha(x,y+\lambda^0(x))={\rm Im}T_\alpha(x,y)+\partial_y{\rm Im}T_\alpha(x,y)\lambda^0(x)\,.
\ee
This clearly induces the following local Peccei-Quinn symmetry for the imaginary parts of the 3d complex scalars:
\be\label{gaugediff}
{\rm Im}T_\alpha(x)\longrightarrow {\rm Im}T_\alpha(x)+M_\alpha\lambda^0(x)\,,
\ee
where it is manifest that the $M_\alpha$'s play the role of the charges of the corresponding scalars under this symmetry. The latter is spontaneously broken by $\langle{\rm Im}T_\alpha\rangle$ and therefore an equivalent, gauge fixed description of the same theory is possible, in terms of a massive vector field: The gauge boson $A^0(x)$ acquires mass by the Higgs mechanism, i.e.~by eating up the Goldstone boson ${\rm Im}T_\alpha(x)$.

Since we have supersymmetrically reduced a four-dimensional $\cN=1$ theory, the resulting action will be a $\cN=2$ 
supergravity theory in three dimensions. Therefore our aim is now to recast the action \eqref{3dkineticT} in the canonical way for such a theory. 
Let us start by recalling the general form of an 3d, $\cN=2$ action in terms of chiral multiplets only, which we call $T_\Sigma$. The 
purely bosonic part of the action is given by \cite{deWit:2003ja,Berg:2002es,Hohm:2004rc}
\bea\label{kinetic_lin_gen_1D}
 S^{(3)}_{\cN=2} = -\int\tfrac{1}{2}R_3 *_3\mathbf{1} 
  +  K_{T_\Lambda \bar T_{\bar\Sigma}}\, 
  \nabla T_{\Lambda}\wedge *_3 \overline{\nabla T_{\Sigma}} +\tfrac12 \Theta_{IJ} A^{I} \wedge F^{J} + V *_3 \mathbf{1} \; ,
\eea 
where $\Theta_{IJ}$ is a symmetric, constant ``embedding tensor'', giving rise to a Chern-Simons term for some non-dynamical vector fields $A^I$, such that $F^I=dA^I$, without kinetic term. The covariant derivatives in this action are taking  the form
\beq \label{DMTheta}
   \nabla T_\Sigma = d T_\Sigma + \Theta_{IJ}X^I_\Sigma \, A^{J} \;,
\eeq
where $X^{I}=X^{I}_\Lambda \partial_{T_\Lambda}+\bar X^{I}_{\bar\Lambda} \partial_{\bar T_{\bar\Lambda}}$ are the Killing vectors generating a subgroup of 
isometries of the K\"ahler manifold with coordinates $T_\Lambda$ which are gauged by $A^I$. 
Such gaugings generate non-trivial D-terms given by momentum maps $D^I$, which fulfill the following relation:
\beq \label{3dDterm}
   i \partial_{T_\Lambda} D^I = K_{T_\Lambda \bar T_{\bar\Sigma}} \bar X^{I}_{\bar\Sigma}\ ,
\eeq
a condition known for the four-dimensional D-terms. The scalar potential in \eqref{kinetic_lin_gen_1D} is:
\be \label{3dPotential}
   V =K^{T_\Lambda\bar T_{\bar\Sigma}}\partial_{T_\Lambda}\cT\partial_{\bar T_{\bar\Sigma}}\cT - \cT^2  
     +e^K \big(K^{T_\Lambda\bar T_{\bar\Sigma}} \cD_{T_\Lambda} W \overline{\cD_{T_\Sigma} W} - 4|W|^2\big) \; ,  
\ee 
where $\cD_{T_\Lambda}=\partial_{T_\Lambda}+K_ {T_\Lambda}$ is the K\"ahler covariant derivative. The scalar potential is given in terms of two functions which, together with the K\"ahler potential $K$, specify completely the 3d, $\cN=2$ action: The holomorphic superpotential $W(T)$, and the D-term potential $\cT(T,\bar T)$, which depends on the momentum maps as
\be\label{cTPotential}
\cT=-\cfrac12D^I\Theta_{IJ}D^J\, ,
\ee
and requires the specification of the embedding tensor $\Theta_{IJ}$.

Let us now specify the characteristic data of the 3d, $\cN=2$ gauged supergravity obtained by 
dimensionally reducing on $S^1$ a 4d, $\cN=1$ theory (eq. \eqref{4dkineticphi}) with 
chiral fields $T_\alpha$ with shift symmetries and no vectors.
In this situation, the indices  $\Lambda$ and $I$ appearing in  \eqref{kinetic_lin_gen_1D} have the same nature and take value in the set $\{0,\alpha\}$, where $0$ corresponds to the new chiral field arising after the reduction, the radion. The superpotential $W=W_{\rm 4d}$ can only depend on the scalars which do not have shift symmetry. 
Moreover, the shift symmetries for Im$T_\alpha$ give rise to isometries of the corresponding moduli space, which are generated by the subset of Killing vectors 
\be\label{ShiftKilling}
X^\Lambda_\Sigma=-2i\,\delta^\Lambda_\Sigma\,.
\ee
Therefore, by inserting \eqref{ShiftKilling} in \eqref{DMTheta} and comparing it with \eqref{InvDerivat}, we are led to identify circle fluxes
with off-diagonal components of the constant embedding tensor, i.e.
\be\label{identificationTX}
\Theta_{\alpha0}\equiv- \frac{M_\alpha}{2} \,.
\ee
In this particular case these are the only non-trivial components of the constant embedding tensor. Thus, being the latter of rank one, the shift symmetry of only one of the $T_\alpha$'s is effectively gauged. The K\"ahler potential in \eqref{kinetic_lin_gen_1D} is related to the kinetic potential \eqref{KineticPotential} via a Legendre transformation with $R$ and Re$T_0$ conjugate variables. Thus we find
\bea\label{3dKaehlerPotential}
K(T_\Sigma+\bar T_{\bar\Sigma})&=&-\log(\RE T_0)+K^{\rm 4d}(T_\alpha +\bar T_{\bar\alpha})\,,\nn\\
\R T_0 &=& \frac{1}{R} \,.
\eea
Note that $K_{T_0 \bar T_{\bar\alpha}}=0$ and $K_{T_\alpha \bar T_{\bar\beta}}=\tilde K_{T_\alpha \bar T_{\bar\beta}}$. Thus the kinetic terms for the scalars $T_\alpha$ in \eqref{kinetic_lin_gen_1D} correctly reproduce the ones in the action \eqref{3dkineticT}. Moreover, we observe that for such a simple K\"ahler potential, a no-scale property holds for the field $T_0$ alone, namely:
\be\label{noscaleT0}
K^{T_0 \bar T_0} K_{T_0}K_{\bar T_0}=1\,.
\ee 
Inserting \eqref{noscaleT0} in \eqref{3dPotential}, and using that $W$ is independent of $T_0$ at this point, we recover the standard scalar potential of 4d, $\cN=1$ supergravity\footnote{In 4 non-compact dimensions, the fluxes \eqref{fluxImT} must vanish in order not to break Poincar\'e invariance of the vacuum.}, with the factor of $3$ in front of $|W|^2$. 

It is easy to see that the D-term potential, which in the present case becomes 
\be\label{DTermPotential}
\cT=\cfrac12D^0 M_\alpha D^\alpha\,,
\ee
gives rise exactly to the scalar potential in \eqref{3dkineticT}. Indeed, by inserting \eqref{ShiftKilling} in \eqref{3dDterm}, one finds
\be\label{MomentumMaps}
D^\Lambda=2K_{T_\Lambda}\,.
\ee
Therefore, by inserting \eqref{DTermPotential} in \eqref{3dPotential} and using \eqref{MomentumMaps}, \eqref{3dKaehlerPotential}, one gets
\be
K^{T_\alpha\bar T_{\bar\beta}}\partial_{T_\alpha}\cT\partial_{\bar T_{\bar\beta}}\cT=R^2\,\tilde K_{T_\alpha \bar T_{\bar\beta}}M_\alpha M_\beta\,.
\ee
Moreover, the contribution to the 3d scalar potential of the other two $\cT$-dependent terms in \eqref{3dPotential} exactly cancel, i.e. 
\be
K^{T_0\bar T_0}\partial_{T_0}\cT\partial_{\bar T_0}\cT - \cT^2 =0\,.
\ee
In order to recover the other terms in the action  \eqref{3dkineticT}, we have to systematically perform  the dualization which brings \eqref{kinetic_lin_gen_1D} to the dual action written in terms of the chiral multiplets $T_\alpha$ and the vector multiplet $(R,A^0)$. As we have just seen, at the level of the K\"ahler potential, such a procedure simply amounts to a Legendre transformation. The relevant terms of the action \eqref{kinetic_lin_gen_1D} are:
\bea\label{ChiralAction}
&&-  K_{T_0 \bar T_0}\, 
  d \,{\rm Re}T_0\wedge *_3  d\, {\rm Re}T_0 -  K_{T_0 \bar T_0}\, 
  (d\, {\rm Im}T_0-2\Theta_{0\alpha}A^\alpha)\wedge *_3  (d\, {\rm Im}T_0-2\Theta_{0\beta}A^\beta)\nn\\ && -  \Theta_{0\alpha} A^{\alpha} \wedge F^{0}\,.
\eea
Note that the action \eqref{ChiralAction} is invariant under the gauge transformation:
\bea\label{GaugeStueck}
{\rm Im}T_0&\longrightarrow&{\rm Im}T_0+2 \Theta_{0\alpha}\lambda^\alpha(x)\,,\nn\\
A^\alpha&\longrightarrow&A^\alpha+d\,\lambda^\alpha\,,
\eea
which, in contrast to the one in \eqref{gaugediff} for the $0$ index, has no geometrical origin.
Let us now use the following dualization algorithm:
\begin{enumerate}
\item Fix the gauge \eqref{GaugeStueck} by setting ${\rm Im}T_0=0$: Hence  $\Theta_{0\alpha}A^\alpha$ becomes a massive vector by  the Higgs 
mechanism.
\item Treat  $\Theta_{0\alpha}A^\alpha$ as a Lagrange multiplier and eliminate it by setting it to its on-shell value.
\end{enumerate}
The equation of motion for $\Theta_{0\alpha}A^\alpha$ reads:
\be\label{EqOfMotion}
\Theta_{0\alpha}A^\alpha=\cfrac18 K^{T_0\bar T_0}*_3 F^0\,
\ee
where we have used that $*_3^{\phantom{l}2}=-1$ with Minkowski signature $(-,+,+)$. Inserting \eqref{EqOfMotion} back into \eqref{ChiralAction} we get exactly the kinetic term for the vector multiplet $(A^0,R)$ in \eqref{3dkineticT}, which now makes $A^0$ a dynamical vector field. In the derivation we have used \eqref{KineticPotential}, which in turn implies Re$T_0=R^{-1}$ and the relation $K^{T_0\bar T_0}=-4\tilde K_{RR}$. Note that no Chern-Simons term survives after the dualization, in the action \eqref{3dkineticT}.

To summarize, once we turn on ${\rm d}({\rm Im}T_\alpha)$ fluxes (allowing a dependence of the imaginary parts of the 4d scalars on the compact coordinate), the corresponding 3d scalars will transform under a local Peccei-Quinn symmetry with charges equal to the flux quanta. From the 4d perspective such gauging is a consequence of the 1d diffeomorphisms along the $S^1$, while from the 3d perspective it is induced by non-trivial Chern-Simons couplings. As remarked above, such symmetry is spontaneously broken and its gauge boson, which is the Kaluza-Klein vector $A^0$, can acquire mass by an ``affine'' Higgs mechanism. Nevertheless, the 3d vector multiplet $(A^0,R)$ can still be dualized to the 3d chiral multiplet $T_0$ since, thanks to the non-trivial embedding tensor $\Theta$, the field $A^0$ will still appear in the dual action as a non-dynamical vector, i.e.~without kinetic term.

One can actually think of yet another, much more subtle contribution to the 3d landscape, namely the one related to the $\Theta_{00}$ term of the constant embedding tensor. In analogy to the fluxes above, this is induced by a non-trivial dependence of the purely 3d field Im$T_0$ on the parameter $y$. This clearly implies an additional gauging for $T_0$, which will now be charged also under the $S^1$-diffeomorphisms. One therefore expects new terms in the scalar potential, dependent on this new charge. These are expected to arise from the reduction of the 4d Ricci scalar $R_4$ in the presence of this more involved geometry. We will not attempt to perform such a reduction here, but rather we will tackle this problem in the M-theory context in section \ref{stringinterpret}, where the structure of the internal elliptic fibration allows to introduce these new fluxes, and at the same time suggests a way to reabsorb them by
a suitable field redefinition.

\subsection{Inclusion of 4d vectors and gaugings}\label{4dVectors}

Let us now introduce vectors in 4d and perform again the fluxed compactification of the previous section. 
We will need this extension in order to generalize non-perturbative contributions to the superpotential 
in the case there are circle fluxes which gauge the shift symmetry of $T_0$. Since the reduction is now much more involved, 
we give here only the result in terms of the 3d, $\cN=2$ action \eqref{kinetic_lin_gen_1D}, with all the 
fields dualized to complex scalars, leaving to appendix \ref{Dualization4dVect} the details of the 
dualization to the vector multiplet formalism.  For the discussion of such reductions without 
circle fluxes, see also \cite{Haack:1999zv,Berg:2002es,Grimm:2010ks,Grimm:2011tb}

For simplicity we will focus on an $SU(N)$ gauge theory. In string constructions such a theory can arise from
stacks of D-branes leading to a $U(N)=SU(N)\times U(1)$ gauge theory. The overall $U(1)$ will often become massive at the 
Kaluza-Klein scale by geometric St\"uckelberg couplings even in the absence of 
fluxes (see \cite{Grimm:2011tb} for a recent discussion). 
We will therefore neglect it in the following, and only make some additional comments on this $U(1)$ in the 
discussion of instanton corrections. The action for the non-Abelian field strength $F= dA + A \wedge A$ of 
$SU(N)$ is given by
\beq\label{4dnonAb}
S^{(4)}_{F}=- \int \tfrac12 {\rm Re}\,\tau\, {\rm Tr} (F \wedge *_4 F )+ \tfrac12 {\rm Im}\, \tau\,  {\rm Tr} (F\wedge F)\ ,
\eeq
where $\tau$ is the gauge coupling function which depends holomorphically on the complex scalar fields. 
As before we will only consider effective theories including the fields $T_\alpha$, and do not allow for charged 
fundamental matter.

In the string 
compactifications the $SU(N)$ can appear in a broken phase due to Abelian gaugings of the $T_\alpha$ arising from background fluxes.
More precisely, the ordinary derivatives in \eqref{4dkineticphi} of the $T_\alpha$ are replaced by 
covariant derivatives
\beq
   \nabla T_\alpha = dT_\alpha + {\rm  Tr} (f_\alpha A)\ , 
\eeq
where $f_\alpha$ is a constant matrix in the adjoint of $SU(N)$.
 
For generic gaugings, provided $n_s\geq N-1$, the $SU(N)$ is completely broken to the Cartan subgroup $U(1)^{N-1}$ with Abelian vectors $A^i$. 
The $\cN=1$, 4d action \eqref{4dkineticphi} then becomes
\bea\label{4dkineticphiVect}
S^{(4)}&=&- \int \tfrac12\,R_4\, *_4{\bf{1}}+ K^{\rm 4d}_{T_\alpha \bar T_{\bar\beta}} \nabla T_\alpha \wedge *_4\overline{ \nabla T_\beta}+\tfrac12 {\rm Re}\,\tau_{ij}F^i\wedge *_4F^j+ \tfrac12 {\rm Im}\, \tau_{ij}F^i\wedge F^j\nn\\
&&\qquad + (V_{\rm F} + V_{\rm D})*_4{\bf 1} + \cL_{\cR^2} \,,
\eea
where $\cL_{\cR^2}$ is the 
higher curvature term given in \eqref{def-cLR2}. 
Since this theory arises from a broken non-Abelian group the classical gauge coupling functions $\tau_{ij}$
are determined from the $\tau$ in \eqref{4dnonAb} by
 \beq 
       \tau_{ij}=C_{ij}\tau\ ,
\eeq       
with $C_{ij}$ obtained from the trace of two Cartan generators in a chosen basis. The non-Abelian gauge symmetry is
broken by gaugings of the form
\be\label{gauging4D}
 \nabla T_\alpha = d T_\alpha +  X_{i\alpha} A^i\ ,\qquad\qquad X_{i\alpha}={\rm  Tr}(f_\alpha\cG_i)\,,
\ee
where $\cG_i$ are the chosen Cartan generators for $SU(N)$ and $X_{i}=X_{i\alpha} \partial_{T_\alpha}+\bar X_{i\bar\alpha} \partial_{\bar T_{\bar\alpha}}$ become the set of Killing vectors generating a subgroup of 
isometries of the K\"ahler manifold with coordinates $T_\alpha$ which are gauged by $A^i$.
Such gaugings induce a non-trivial D-term potential
\be
V_D=\tfrac12({\rm Re}\,\tau)^{ij}D_i D_j\,,
\ee
where $\R \tau^{ij}$ is the inverse of $\R \tau_{ij}$, and the D-terms satisfy the relation
\be\label{4dDterm}
 i \partial_{T_\alpha} D_i = K_{T_\alpha \bar T_{\bar\beta}} \bar X_{i\bar\beta}\,.
\ee
Note that we will restrict to the case of gauged shift symmetries \eqref{shift-symmetry} with constant $X_{i\alpha}$ 
such that \eqref{4dDterm} is readily integrated. The integration constant in believed to be 
zero in string compactifications.

In the next step we will restrict to a leading gauge coupling function $\tau$ linear in the complex 
scalars $T_\alpha$. This is analogous to \eqref{simple_sigma}, and in summary we have 
\beq \label{sigma_tau}
   \tau = \tfrac12 C^\alpha T_\alpha \ , \qquad  \sigma = \tfrac18 k^\alpha T_\alpha\ ,
\eeq
with $C^\alpha$ and $k^\alpha$ being constant vectors. 
We will assume that these two couplings are well-defined along $S^1$, which implies that  the field strength of their imaginary parts cannot have any flux.
In other words, when imposed on the shift symmetries, we have the restriction 
\beq \label{CQ=0}
   C^\alpha Q^\Sigma_\alpha = 0 \ ,
\eeq
where $Q^\Sigma_\alpha$ is parameterizing the shift symmetries as in \eqref{shift-symmetry}. 
This is in complete analogy to \eqref{ktheta} for the $\cR \wedge \cR$-term. Using the 
definition of the fluxes $M_\alpha$ in \eqref{fluxImT} we find that 
\be\label{CQ=0M}
C^\alpha M_\alpha =0\,.
\ee
Note that gauging the shift symmetries parameterized by $Q_\alpha^\Sigma$ amounts to 
specifying constant Killing vectors $X_{i\alpha}$ picked among the set of $Q$'s. 
From the conditions \eqref{ktheta} and \eqref{CQ=0} one thus infers
\be\label{Assumption}
C^\alpha X_{i\alpha}=0\,,\;\qquad \qquad\; k^\alpha X_{i\alpha}=0\qquad\qquad\forall i\,.
\ee
These conditions assure that in the 4d theory $\tau$ and $\sigma$ are not gauged by any of the $A^i$, 
so that the  action \eqref{4dkineticphiVect} is classically gauge invariant. 


We now aim to reduce this more general 4d theory down to 3 dimensions on $S^1$ with the background 
metric given in \eqref{backmetric} and fluxes \eqref{fluxImT}. 
We take again the metric fluctuation of the form \eqref{metricfluct}, while the 4d vector fields split as follows:
\bea\label{4d3dvectors}
A^i_p&=&\left(A^i_\mu-A^0_\mu\zeta^i,\zeta^i\right)\,,\\
\langle A^i\rangle&=&\left(0,\langle\zeta^i\rangle\right)(y)\,,\label{BackgroundZeta}
\eea
where by abuse of notation we denote also the 3d vectors by $A^i$, and $\zeta^i$ are 3d real scalars. Therefore, in addition to the vector multiplet of the graviphoton, we will now have new 3d vector multiplets, with bosonic content given by:
\be \label{3Dred_vec}
\left(A^i, \xi^i\right)\qquad\qquad\xi^i\equiv R\zeta^i\,.
\ee
Note that non-trivial background values \eqref{BackgroundZeta} for the 4d vectors are in general allowed along the $S^1$. Due to the large gauge transformations of the 4d vectors, these vevs undergo integral shifts
\be\label{LargeGauge}
\langle\zeta^i\rangle\longrightarrow\langle\zeta^i\rangle+\ell^i\,,\quad\qquad\ell^i\in\mathbb{Z}\,.
\ee 
In addition, the definition  \eqref{fluxImT} for the fluxes slightly 
changes here due to the presence of 4d gaugings which make \eqref{fluxImT} no longer invariant under large gauge transformations. 
This can be cured by replacing the simple derivative with the covariant one as
\be\label{GaugeInvCircleFlux}
M_\alpha= \int_{S^1}\langle\nabla \I T_\alpha\rangle\,,
\ee
which is now a consistent, gauge invariant definition.
Let us observe that, if we want to keep Poincar\'e invariance in three dimensions, no fluxes 
can be introduced which arise from the 4d vectors.\footnote{One could only think of integrating their field strengths on the compact direction, but this would clearly generate a Lorentz-breaking flux in 3d.} 

As we noted above, in three dimensions we can eliminate all dynamical vectors in favor of scalars to bring the action into the form \eqref{kinetic_lin_gen_1D}. In order to do that for the vectors in $(A^i,\xi^i)$ and $(A^0,R)$ we introduce the dual complex coordinates
\bea \label{def-Ti-T0}
   T_i & = & - \frac{2}{R} \, \R \tau_{ij}\, \xi^i   + i \tilde \zeta_i \label{TiFields}  \ ,\\
   T_0 &=& \frac{1}{R} +  \frac{1}{R^2} \, \R \tau_{ij}\, \xi^i \xi^i     +  i\tilde \zeta_0 \label{T0Field}\ .
\eea
Here $(\tilde \zeta_i,\tilde \zeta_0)$ are the dual real scalars which carry the dynamical degrees of freedom of the vectors $(A^i,A^0)$.
Note that $\R T_i$ and $\R T_0$ are connected with $R,\xi^i$ via a Legendre transform as we discuss in appendix \ref{Dualization4dVect}. The
resulting K\"ahler potential is given by 
\beq\label{KaehlerPotVect}
   K(T_{\Sigma}+\bar T_{\bar\Sigma}) = - \log\Big(\RE T_0 - \tfrac14 \R\, \tau^{ij} \RE T_i  \RE T_j  \Big)  + K^{4d}(T_\alpha+\bar T_{\bar\alpha})\ ,
\eeq
where $T_\Sigma = (T_0,T_i,T_\alpha)$. As a simple check, one notes that
the expression for $K$ and $T_0$ reduces to \eqref{3dKaehlerPotential}  in the absence of vectors, i.e.~when $T_i=0$. Furthermore, let us observe that this K\"ahler potential no longer fulfills the no-scale property \eqref{noscaleT0}. The latter is indeed violated by the presence of vectors in the following way:
\be
K^{T_0 \bar T_0} K_{T_0}K_{\bar T_0}=1+\frac{2{\rm Re}\,\tau_{ij}\xi^i\xi^j}{R}\,.
\ee  
It is then easy to see that the K\"ahler potential \eqref{KaehlerPotVect} satisfies the new no-scale property:
\be
K^{T_I \bar T_{\bar J}} K_{T_I}K_{\bar T_{\bar J}}=1\,,
\ee
where $I=\{0,i\}$.

After integrating over $S^1$, performing the Weyl rescaling, and dualizing all the vector multiplets to chiral multiplets, the 3d action 
can be brought into the form \eqref{kinetic_lin_gen_1D}, with an additional term given by the reduction of the higher derivative correction 
$\cL_{\cR^2}$. In the following we will not need the explicit form of this higher curvature terms. However, as a crucial result of the 
reduction including the fluxes $M_\alpha$ and gaugings $X_{\alpha i}$ we find a modified 
covariant derivative of the form:
\be \label{covariantTLambda}
  \nabla T_\Lambda = d T_\Lambda -2 i\,\Theta_{\Lambda\Sigma}A^{\Sigma} \;,
\ee
where generally all vectors $A^\Lambda = (A^0,A^i,A^\alpha)$ can appear in the gauging. 
Explictly the gaugings are determined by
\be\label{ThetaX0}
 \Theta_{0\alpha} = -\frac{M_\alpha}{2}\, , \qquad \Theta_{i\alpha}=\frac{i}{2}\,X_{i\alpha}\,,\qquad\Theta_{\alpha\beta}=\Theta_{00}=\Theta_{0i}=\Theta_{ij}=0\, ,
\ee
which generalizes \eqref{identificationTX}.
The scalar potential is given by eq. \eqref{3dPotential} with $W = W_{\rm 4d}$ 
and a D-term potential 
\be\label{ExplicitDtermPot}
\cT=-\cfrac12D^\Lambda\Theta_{\Lambda\Sigma}D^\Sigma=-\frac{K_{T_\alpha}\left(M_\alpha+\frac{i}{2}X_{i\alpha}{\rm Re}\,\tau^{ij}{\rm Re}\,T_j\right)}{{\rm Re}\,T_0-\tfrac14{\rm Re}\,\tau^{ij}{\rm Re}\,T_i{\rm Re}\,T_j}\,,
\ee
with 3d D-terms $D^\Lambda$ still given by eq.~\eqref{MomentumMaps}. 
In the last equality we used \eqref{KaehlerPotVect} and \eqref{ThetaX0}. The explicit computation of the 
scalar potential can be found in appendix \ref{Dualization4dVect}, where it is shown to coincide with 
the one obtained by dimensionally reducing the 4d action \eqref{4dkineticphiVect}.

\subsection{Taub-NUT instantons and the stabilization of the radion}\label{StabRadion}

In this section we will include a particular set of supersymmetric corrections to 
the three-dimensional effective theory.
In particular we will argue that four-dimensional gravitational instantons, more precisely Taub-NUT geometries, 
will appear as corrections in the three-dimensional 
superpotential $W$.
This will also provide us with a 
mechanism to stabilize the radion field $r$, or rather the combination $T_0$ introduced in \eqref{def-Ti-T0}. 
In the presence of a four-dimensional superpotential $W_0 = W_{4d}$, and the 
absence of fluxes $M_\alpha$ along the $S^1$ this stabilization mechanism will 
be completely universal. 
The generalization including fluxes will be significantly more complicated and 
discussed separately in section~\ref{FluxSuperpotential}.

Let us first assume that the three-dimensional superpotential is independent of the scalar $T_0$
parameterizing the size of the fourth dimension. In this case the superpotential is 
of the form \cite{Seiberg:1996nz,Katz:1996th} 
\beq \label{3dsuper_noTN}
   W = W_{4d}(T_\alpha) + \sum_i A_i e^{-2 \pi T_i}  + \tilde A e^{-2 \pi \tau} e^{2\pi \sum_i a_i T_i}
\eeq
where $a_i$ are the Dynkin labels of the gauge group in \eqref{4dnonAb}, e.g.~$a_i=1$ for $SU(N)$. 
Inserting this into the three-dimensional scalar potential \eqref{3dPotential} using the K\"ahler 
potential \eqref{KaehlerPotVect} one finds that the radion $r$ has no local minimum at finite $r$.
If in the moduli stabilization the negative term in \eqref{3dPotential} 
is canceled, $r$ tends to minimize the potential in the limit $r \rightarrow \infty$. 
This is due to the fact 
that the scalar potential \eqref{3dPotential} admits an overall factor $e^{K}=\frac{e^{K^{4d}}}{r^2}$,
where $K^{4d}$ is the four-dimensional K\"ahler potential. Clearly, this limit 
corresponds to the decompactification limit to four space-time dimensions. 

Let us now ask if one can obtain a local minimum at finite $r$ by including further corrections to 
the 3d superpotential. Firstly, note that there cannot be any perturbative corrections in $T_0$  
to $W$. This is due to the fact that $\I T_0$ has a perturbative shift symmetry, since it arises from dualizing 
a 3-dimensional vector $A^0$. The immediate possibility is a new non-perturbative correction of the form 
\beq \label{Taub-NUT_super}
  W(T_0) =  \cA \, e^{-2 \pi\, T_0}\ , 
\eeq
where  $\cA$ is some $T_0$-independent pre-factor which will be discussed in more detail below. 
The first check on such a superpotential is provided by gauge invariance. However, since 
we focus on the fluxless reduction  $M_\alpha=0$ this is trivially the case since $T_0$ is not charged. 
In addition one has to check the zero mode condition, and hence identify the source 
of the correction \eqref{Taub-NUT_super}. Since $T_0$ parameterizes degrees of freedom of 
the 4d metric, the superpotential is expected to be induced by a four-dimensional gravitational instanton. 
Let us note that the
gravitational instanton of interest is a Euclidean Taub-NUT space and its AdS generalizations.\footnote{We Wick-rotate the time direction because we consider 
Taub-NUT instantons which will later admit gauge instantons.} 

To motive that \eqref{Taub-NUT_super} arises from a Taub-NUT gravitational instanton 
let us evaluate the four-dimensional action.
We first summarize the key properties of the Taub-NUT gravitational solution. 
The Taub-NUT is a special case of an ALF space (asymptotically locally flat space). It is a non-compact
manifold which is everywhere a fibration of $S^1$ over $\bbR^3$, apart of a single point, the nut, at which 
the circle shrinks to zero size. Its asymptotic boundary $\partial {\rm TN}$ is topologically a three-sphere $S^3$ realized as
an Hopf fibration of $S^1$ over $S^2$ with first Chern number $1$. 
The metric of Taub-NUT is well-known \cite{Eguchi:1980jx,Gibbons:1994cg} and can be written as follows
\be\label{MetricTaubNUT}
ds^2=\left(\frac{1}{\lambda^2}+\frac{1}{\rho}\right)\left(d\rho^2+\rho^2\left(d\theta^2+\sin^2\theta d\phi^2\right)\right)
+\frac{1}{\frac{1}{\lambda^2}+\frac{1}{\rho}}\left(d\psi+\cos\theta d\phi\right)^2\,,
\ee
where $\rho\in[0,\infty)$ parameterizes the radial distance from the nut, $\theta\in[0,\pi)$, and $\phi\in[0,2\pi)$ are the angular coordinates of the $S^2$ at infinity of the base ($\rho\to\infty$), and $\psi\in[0,4\pi)$ parameterizes the fiber $S^1$. The real number $\lambda$ represents the asymptotic radius of the fiber and therefore it is proportional to the expectation value of the radion field on this Tub-NUT background, i.e.~$\lambda=\langle r\rangle/4\pi$. In the limit $\lambda\to\infty$, the metric \eqref{MetricTaubNUT} reduces to the flat metric on $\bbR^4$. 
Using \eqref{MetricTaubNUT} one finds that the curvature form on this space is anti-self-dual, i.e.~$* \cR_4 =- \cR_4$.
The topology of TN is that of $\bbR^4$ and yields a first Pontrjagin number 
\beq \label{Pontryagin}
   p = - \frac{1}{2}\int_{\rm TN} \text{Tr}(\cR_4 \wedge \cR_4) = 2\ .
\eeq
Due to the non-compactness of the Taub-NUT space the evaluation of the 
action is non-trivial, since it has to be regularized. This was first done 
in ref.~\cite{Gibbons:1979nf} by subtracting the action of Euclidian flat space, despite 
the fact that the boundaries of these two non-compact spaces do not match.
 The 
resulting action was found to be 
\beq \label{classical_TN_action}
  S^{(4)}_{\rm TN} = - \frac{4 \pi n^2}{G}   = - \frac{\langle r\rangle^2 M_p^2}{2} = - 2\pi \langle r\rangle^2 \ , 
\eeq
where we have expressed the so-called nut-charge $n =  \langle r\rangle/8 \pi$ in terms of the circumference 
$\langle r\rangle$ of the periodic Euclidian direction at infinity. In order to evaluate \eqref{classical_TN_action}
we have used that our action \eqref{4dkineticphi} is negative, and the reduced Planck mass square $M^2_p$ is $4\pi $ in the conventions of \cite{Grimm:2011tb}. 
The action \eqref{classical_TN_action} was later confirmed using the AdS-CFT prescription of \cite{Balasubramanian:1999re} to 
regularize the gravitational action of Taub-NUT-AdS$_4$ in \cite{Emparan:1999pm}. 
For Taub-NUT-AdS$_4$ the action is given by  
\beq \label{TN-AdS_action}
   S^{(4)}_{\rm TN-AdS} = - \frac{4 \pi n^2}{G} \Big(1 - \frac{2 n^2}{l^2} \Big)
\eeq
where $l$ is the AdS-radius. This indeed yields \eqref{classical_TN_action} when sending $l$ to infinity.
One realizes that in the absence of vectors the action $S^{(4)}_{\rm TN}$ in \eqref{classical_TN_action} agrees with the 
real part of $T_0$, $\R T_0 = r^2$. It would be interesting to perform the reduction around an 
AdS-background and match the action \eqref{TN-AdS_action}.

Let us include the higher curvature terms \eqref{def-cLR2} and denote 
this corrected action by $\tilde S^{(4)}_{\rm TN}$.
One then 
obtains by using the anti-self-duality of $\cR$ and \eqref{Pontryagin} that the Taub-NUT gravitational instanton 
contributes in the 3d effective theory with an exponential $e^{\tilde S^{(4)}_{\rm TN}}$. 
Hence, one finds that the correction to the superpotential is of the form 
\beq\label{TaubNUTSigma}
  W(T_0) = \cA e^{- 8\pi \sigma} e^{-2\pi T_0}\ . 
\eeq

A by now well-known proposal to stabilize moduli using instanton corrections 
was given by KKLT \cite{Kachru:2003aw}. Here we suggest that the correction \eqref{TaubNUTSigma}
precisely allows us to generate a new vacuum at finite $r$. In order to do that
one can proceed in various ways, depending on the scale at which the 
fields $T_\alpha$ are stabilized. We discuss two scenarios in turn.

\begin{figure}[h!] 
  \centering
      \includegraphics[width=13cm]{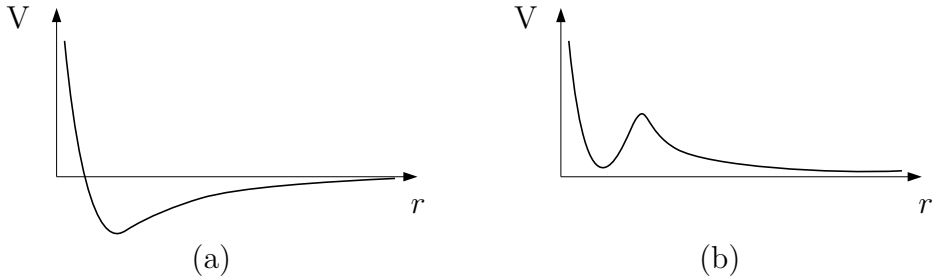}
  \caption{Scalar potential for the radion $r$ before (a) and after (b) the up-lift.} \label{KKLT_pot}
  \begin{picture}(0,0)
 \put(-047,60){$r$}
  \put(145,60){$r$}
 \put(-200,130){V}
  \put(-8,130){V}
  \put(-130,39){(a)}
  \put(62,39){(b)}
\end{picture}
\end{figure}

Let us first consider the case that all fields $T_\alpha$ have been supersymmetrically fixed 
in 4d solving $D_{T_\alpha } W_{\rm 4d} =0$ with the  $T_\alpha$ appearing in the superpotential.  
If the masses of such stabilized $T_\alpha$ 
are sufficiently large they can be integrated out from the 4d effective theory.
In the vacuum this can lead to four-dimensional $W_{\rm 4d} = W_0$ 
in \eqref{3dsuper_noTN}, where $W_0$ is a small
constant. For example, in a Type IIB compactification $W_0$ can be determined by a 
4d flux superpotential as in \cite{Dasgupta:1999ss,Giddings:2001yu,Kachru:2003aw}. 
Clearly, this implies that the 4d effective theory 
has a scalar potential
\beq \label{AdS_4_pot}
    V_{\rm 4d} =  - 3 e^{K^{\rm 4d}} |W_0|^2\ ,
\eeq
with negative cosmological constant. The supersymmetric solutions are 
AdS$_4$ vacua.
Considering this effective theory compactified on a circle, 
the three-dimensional superpotential is expected to be of the form 
\beq \label{W0+expT}
   W = W_0 +  \cA \, e^{-2 \pi T_0}\ ,
\eeq
where we note that the non-perturbative contribution should now arise from
Taub-NUT-AdS$_4$ gravitational instantons.
Together with the K\"ahler potential \eqref{3dKaehlerPotential} one obtains a supersymmetric 
AdS$_3$-vacuum at a value $\langle T_0\rangle$  obtained by solving $D_{T_0}W=0$. This is depicted in 
figure \ref{KKLT_pot}.(a). For sufficiently small $\langle \R T_0 \rangle$ the theory is effectively three-dimensional.

A second possibility is to consider the case where instanton corrections in $T_\alpha$ and $T_0$ 
are included altogether in the 3d effective superpotential. 
This can be the case if the masses of the $T_\alpha$ are of the  
same order as the mass of $T_0$. In the evaluation of the Taub-NUT instanton action one then neglects 
all instanton corrections to the 4d superpotential. If the 4d K\"ahler potential for the $T_\alpha$ is of no-scale type \cite{Cremmer:1983bf},
the scalar potential depending on $W_{\rm 4d}$ will vanish identically even in the presence of a 
non-vanishing $W_{\rm 4d}= W_0$. 
Non-perturbative corrections in the $T_\alpha$, together with the Taub-NUT 
instanton corrections \eqref{TaubNUTSigma} then induce a 3d superpotential, 
$W= W_0 + \sum_\alpha \cB_\alpha e^{ - 2 \pi n_\alpha T_\alpha } + \cA e^{-2 \pi T_0}$, which generically admits 
AdS$_3$ vacua at values $T_0,T_\alpha$ obtained by solving $D_{T_0}W=D_{T_\alpha}W=0$.
As we will see in section \ref{stringinterpret} this possibility is most directly realized in an M-theory compactification 
on Calabi-Yau fourfolds to three dimensions. In this case the $T_0,T_\alpha$ are on the same footing, since 
they measure sizes of divisors. Also the non-perturbative corrections uniformly arise from wrapping 
M5-branes on these divisors, as we discuss in more detail below.

Let us end this section by commenting on a version of this scenario which admits a 3d vacuum with positive cosmological 
constant. Vacua with positive 
cosmological constant, in particular dS$_4$ vacua, are hard to realize within 
a string compactification. A prominent proposal was 
given by KKLT \cite{Kachru:2003aw}, where  anti-D3-branes in a warped throat up-lift 
an AdS$_4$ to a dS$_4$ vacuum. While this suggestion has been disputed over the last 
years, see e.g.~\cite{DeWolfe:2008zy,Bena:2009xk,McGuirk:2009xx,Blaback:2010sj,Dymarsky:2011pm,Burgess:2011rv}, 
let us assume for a moment that such an uplift is possible.
We can imagine two possibilities to link it with the 4d-3d story, which match the two given ways  
to stabilize the $T_0,T_\alpha$ introduced in the last two paragraphs.  Firstly, one could consider a 4d theory in 
which the $T_\alpha$ are sufficiently massive and can be stabilized at a non-supersymmetric dS$_4$ vacuum. 
Both in deriving the effective 3d theory, and the gravitational instanton solution, which is expected 
to be Taub-NUT-dS$_4$ in this case, the fact that the fields $T_\alpha$ settle in a non-supersymmetric vacuum should be 
taken into account. The potential is expected 
to be of the form as given in figure \ref{KKLT_pot}.(b).  The second possibility is the analog of KKLT up-lift in 3d, 
in which the positive cosmological constant is treated as a perturbation of a supersymmetric vacuum.
The string realization would be an M-theory compactification with anti-M2-branes localized in 
a warped throat. Clearly it would be interesting to make these proposals more concrete. 
In particular, this would open the possibility to study tunnelings from an effectively 3d theory with a dS$_3$ vacuum
to an effectively 4d theory  with a cosmologically relevant vacuum. The cosmological implications 
of such transdimensional tunnelings have been discussed in  \cite{Linde:1988yp,Giddings:2004vr,BlancoPillado:2009di,Carroll:2009dn,Graham:2010hh}.

\subsection{Superpotentials in the presence of fluxes}\label{FluxSuperpotential}

In this subsection we want to generalize the discussion of the superpotential to the 
case where vector fields and circle fluxes $M_\alpha$ are included in the reduction as 
in subsection~\ref{4dVectors}. Recall that in this more general situation 
new gaugings are present. Using \eqref{ThetaX0} one finds that under the 3d $U(1)$ gauge 
transformations for $A^\alpha$ the scalars have to shift as
\bea
   T_0 & \longrightarrow & T_0-i\,M_\alpha\lambda^\alpha \, ,\label{GaugingT0} \\
   T_i & \longrightarrow & T_i- X_{i\alpha}\lambda^\alpha \,,\label{GaugingTi} 
\eea
where $\lambda^\alpha(x)$ are the local gauge shifts.
These transformations ensure the invariance of the covariant derivatives \eqref{covariantTLambda}. 
One realizes that the first transformation \eqref{GaugingT0} implies that 
a superpotential $W=\cA e^{- 2 \pi T_0}$ is no longer gauge invariant for a 
constant prefactor $\cA$ due to the non-vanishing fluxes $M_\alpha$. 
In the following we want to discuss a simple way to restore gauge invariance by 
using a $T_i$ dependent prefactor $\cA$ and the transformation \eqref{GaugingTi}. 
This is analog to the situation for fluxed D3-brane instantons in Type IIB and F-theory \cite{Grimm:2011dj}.
In \cite{Grimm:2011dj} the non-trivial cancellation of the gauge transformations arises from the possibility 
to switch on two-form fluxes on the world-volume of the D3-brane instanton wrapped on a four-cycle with 
non-trivial topology. Here we also suggest an interpretation of the `fluxes' 
which have to be included on top of the Taub-NUT geometry. 

To begin with let us first note that any correction to the holomorphic superpotential 
now has to be holomorphic in the complex coordinates $T_0,T_i$ given in \eqref{TiFields} and \eqref{T0Field}.
In particular we realize that the vectors also correct the definition of 
$T_0$ and a simple matching of the Taub-NUT action alone with the $T_0$ is not possible.
This is not surprising, since one now has to include an non-trivial field 
configuration for the 4d vector fields $A^i$ into the discussion.  
We are looking for instanton 
solutions for the gauge fields which have an anti-self-dual field strength 
\beq\label{SelfDualityInstanton}
      \cF = -* \cF \ .
\eeq
This relation should be equally true for the case of $SU(N)$ and $U(1)$ field strengths. 
In the absence of gaugings it is not hard to show that in the case of an anti-self-dual gauge bundle the Einstein 
equations are not modified. This implies that the Taub-NUT geometries 
are still solutions to the equations of motion of the more general action.

We now want to evaluate the 4d Euclidean action including the instanton gauge field 
configuration. As in the previous subsection we will focus on the 
real part of the Euclidean action, and later use holomorphicity of the superpotential 
to find the complete coordinate dependence. Of crucial importance will be 
the kinetic term of the gauge fields\footnote{In this expression we used the convention $M^2_p=4\pi$, as in \eqref{classical_TN_action}.} 
\beq\label{KineticGaugeInstanton}   
   S_{\cF} = -2\pi \int_{\rm TN} \R\, \tau\, \text{Tr} (\cF \wedge * \cF )\quad \rightarrow \quad  S_{\cF} = -2\pi \int_{\rm TN} \R\, \tau\ C_{ij}\,\cF^i \wedge * \cF^j\ , 
\eeq
where $\tau$ is constant in the instanton configuration.
The gauge bundle configuration on Taub-NUT can be constructed for non-Abelian gauge groups 
$U(N)$ as discussed in \cite{Cherkis:2008ip,Witten:2009xu,Cherkis:2009jm,Cherkis:2010bn}. Since we aim to compute corrections 
to the 3d action in the Coulomb phase $U(1)^{N}$ we will 
focus on the $U(1)$ case with vectors $\cA^i$ here.
One first realizes that due to the triviality of the second cohomology $H^2({\rm TN},\bbZ)$ 
of TN one finds that the $U(1)$ field strength configuration is globally exact $\cF^i = d \cA^i$. 
As boundary conditions for the gauge configurations $\cA^i$ we will 
impose a matching with the reduction of 
subsection \ref{4dVectors} and expand
\beq \label{cAexpand}
\cA^i= \langle\zeta^i\rangle \, \Lambda + n^i \, \Lambda\ ,    
\eeq
where $\Lambda$ is defined as 
\be
   \Lambda=\frac{\rho}{4\pi(\rho+\lambda^2)}\,
\left(d\psi + \text{cos} \theta d\varphi \right)\ ,
\ee 
with coordinates as in the metric \eqref{MetricTaubNUT}.
Hence the background gauge field strengths will be
\be\label{LargeGaugeInv}
\cF^i= \left(\langle\zeta^i\rangle+ n^i\right) \, d\Lambda\ ,
\ee
where  $d \Lambda$ is the unique normalizable harmonic two-from on TN.
Using the metric \eqref{MetricTaubNUT} one checks that $* d\Lambda =- d\Lambda$,
such that $\cF^i$ is anti-self-dual. 
Let us comment on the split of \eqref{cAexpand} into $\langle\zeta^i\rangle$ and $n^i$. 
The $\langle\zeta^i\rangle$ parameterize the monodromies (Wilson lines) of the gauge fields $\cA^i$ on the $S^1$ fiber at infinity ($\rho\to\infty$), and 
are all chosen to be in the range $[0,1)$. The $n^i$ are integers and determine the topological sector 
which the instanton configurations $\cA^i$ belong to. They have an intrinsic meaning because 
they make the field strengths in \eqref{LargeGaugeInv} invariant under the large gauge 
transformations \eqref{LargeGauge}.\footnote{This can be viewed as an analog to the 
situation in D-brane configurations where the NS-NS B-field appears with the brane 
fluxes in the gauge-invariant combination $B+F$.} In other words, $n^i$ determines a canonical 
representative in the holonomy class of the configuration $\cA^i$, and can be thought of as the first Chern class of the $i$-th Abelian instanton on TN. Since, as already pointed out, TN has trivial topology, such a class takes values in a ``geometrical'' version of $H^2({\rm TN},\bbZ)$ \cite{Witten:2009xu}. All this has a natural, topological formulation on the manifold which compactifies TN, namely $\bbC\bbP^2$  (see e.g.~\cite{Atiyah:2011fn}). Indeed, we have a natural surjective map
\be
f:{\rm TN}\longrightarrow\bbC\bbP^2\,,\quad\qquad f|_{S^3_\infty}\equiv\pi^{\rm H}:S^3_\infty\longrightarrow S^2\,,
\ee
which fixes TN at finite $\rho$, while sending its infinity to an $S^2$ via the Hopf projection. This two-sphere describing the infinity of TN in its compactification represents (the Poincar\'e dual of) the hyperplane class $\omega$ of $\bbC\bbP^2$, which generates $H^2(\bbC\bbP^2,\bbZ)\simeq\bbZ$. Moreover, one has
\be\label{PullBack}
d\Lambda=f^*(\omega)\,.
\ee

Using \eqref{SelfDualityInstanton}, \eqref{LargeGaugeInv} and \eqref{PullBack} we can easily evaluate \eqref{KineticGaugeInstanton} and we obtain\footnote{We omit the brackets $\langle\rangle$, but we remind that all fields are set to their expectation values on the instanton background under consideration.} \cite{Witten:2009xu}:
\bea \label{action_with_F}
  S^{(4)}_{{\rm TN}+\cF} &=& S^{(4)}_{\rm TN} - 2 \pi \int_{\rm TN} \R\, \tau\ C_{ij}\,\cF^i \wedge * \cF^j  \\
                 &=& -2 \pi r^2 - 2\pi \R \tau \big( n^i n^j C_{ij} + 2 n^i C_{ij} \zeta^j + C_{ij} \zeta^i \zeta^j\big) \nn \\
                 &=& -2 \pi\, \big( \R T_0 - n^i \, \R T_i +  n^i n^j C_{ij}\, \R \tau \big)\ , \nn
\eea
where we have used $\xi^i = R \zeta^i$, and the definitions \eqref{TiFields}, \eqref{T0Field}  of the coordinates $\R T_0$, $\R T_i$.
Note that \eqref{action_with_F} is only a part of the full 4d, $\cN=1$ supergravity action \eqref{4dkineticphiVect} evaluated on this 
instanton background. In fact, one uses that in the background the kinetic terms of the scalars $\R T_\Lambda$ vanish
for their constant vacuum expectation values. The higher curvature terms can be included as in the 
previous subsection and yield an additional term linear in the complex field $\sigma$. 
More subtle are the proper inclusion 
of the imaginary parts of $T_\Lambda$, and the scalar potential. Firstly, $\I T_0$, $\I T_i$ only arise after dualizing 
components of the 4d metric and 4d vectors, respectively. Secondly, already in 4d the fields $\I T_\alpha$ appear in gauge covariant 
derivatives. The scalar potential can have a background value \eqref{AdS_4_pot} and hence can also contribute deforming the 
TN into an AdS-Taub-NUT geometry with action \eqref{TN-AdS_action}. Clearly, it would be desirable to 
include these modification in the analysis. In particular, it might be that the gaugings actually modify
the metric solution itself. We will leave the study of the backreaction to future work.

In the following we will take a more basic approach to analyze the superpotential. 
We simply complexify \eqref{action_with_F} such that it can describe 
a correction to the holomorphic superpotential $W$, by replacing 
the real parts of the complex K\"ahler coordinates with their complete expressions.
This leads to the following form of the $T_0$-dependent contribution:
\be\label{FieldDependB}
W(T_0) =  \cC\,e^{-8 \pi \sigma}\, \Big(\sum_{\bar n,n^i} e^{-2 \pi \bar{n}\tau}\, e^{-2\pi C_{ij} n^i n^j \tau +2\pi n^i T_i}\,\Big)\, e^{-2\pi T_0}\,,
\ee
provided the following additional condition holds
\be\label{GaugeInvW}
M_\alpha=-i\,n^iX_{i\alpha}\qquad\forall\alpha\,,
\ee
which assures gauge invariance of \eqref{FieldDependB}. The form of $W(T_0)$ is reminiscent of a 
constraint Jacobi form, as will be also apparent from 
the discussion of the M-theory realization using M5-branes \cite{Witten:1996hc}.\footnote{See also refs.~\cite{Ganor:1996pe,Grimm:2007xm} for related discussions.}

If $N-1$, which is the rank of $SU(N)$, 
is bigger than $n_s$, i.e.~the number of 4d scalars which posses shift symmetry, 
then\footnote{If we suppose the gauge group $SU(N-n_s)$ left unbroken by the gaugings to be further broken, say to its Cartan torus, by other fluxes, we can allow $X_{i\alpha}$ to be an $(N-1)\times n_s$ matrix, as for $n_s> N-1$.} 
\eqref{GaugeInvW} could in principle be satisfied by suitably choosing the $n^i$s. 
However, in general, eq.  \eqref{GaugeInvW} is rather a constraint on the 
fluxes $M_\alpha$, which therefore cannot be completely arbitrary if we 
want gravitational instantons to contribute at all to the 3d superpotential.
 They must be linear combinations of the gaugings appearing in 4d, 
with coefficients equal to the weights $n^i$ appearing in the non-perturbative superpotential. 
Notice that the $\sigma,\tau$-dependent factors in \eqref{FieldDependB} are separately gauge invariant.
Their charges vanish due to conditions  \eqref{kthetaM}, \eqref{CQ=0M} and \eqref{Assumption}. 

Let us conclude this section with an observation. The term in the superpotential \eqref{FieldDependB} 
with coefficient $\bar n$ depends on the holomorphic gauge coupling function $\tau$, which 
in our case is simply given by $\tau=\frac12 C^\alpha T_\alpha$. 
The coefficient $\bar{n}$ 
has an interpretation analogous to the $n^i$s for the central $U(1)$ gauge field 
in $U(N)$. If we restore the non-Abelian phase of the gauge theory by 
switching off the gaugings, we can have a fully non-Abelian $U(N)$ instanton.
More precisely, requiring that the Wilson line background of the instanton 
configuration is trivial, $\bar n$ is an integer instanton number 
given by the second Chern character 
\be\label{nBar}
\bar n=-\frac{1}{2}\int_{\rm TN}{\rm Tr}\,\cF\wedge \cF\,,
\ee 
where $\cF$ is the $U(N)$-valued curvature of the instanton bundle in units of $2\pi$.

In section \ref{stringinterpret} we will provide a
description of the Abelian and non-Abelian instanton bundles which appeared here by means of a M5-brane construction. In the Type IIA weak coupling limit we interpret $\bar n$ and $n^i$ as  wrapping numbers of extended D-brane and NS-brane instantonic sources.


\section{M-theory realization and the F-theory limit}\label{stringinterpret}

In this section we will show how the 
supergravity scenario outlined in section \ref{from4to3} 
can be embedded into an M-theory compactification. 
We first briefly review in subsection \ref{mod_id} the dimensional reduction of M-theory on 
an elliptically fibered Calabi-Yau fourfold, and summarize how the resulting
3d effective theory can be connected to a 4d F-theory compactification. 
The inclusion of four-form fluxes and the discussion of their induced D-term potential 
is given in subsections \ref{Mfluxes} and \ref{Minterpret}. This allows us to make contact with the 
3d gauged supergravity theories obtained in section \ref{from4to3} by fluxed reduction 
on a circle. Finally, in subsection \ref{Taub-NUT_M} we present a microscopic realization of the 
gravitational and gauge instanton corrections introduced in the 4d context in subsections \ref{StabRadion} and \ref{FluxSuperpotential}.
This requires a discussion of M5-brane instantons and their anomalies.

\subsection{M-theory on Calabi-Yau fourfolds and the F-theory limit} \label{mod_id}

We first recall 
the basic steps to link M-theory and Type IIB string theory \cite{Denef:2008wq,Weigand:2010wm}. 
Consider M-theory compactified on a two-torus $T^2$, naming one of the 1-cycles the $A$-cycle, and 
the other 1-cycle the $B$-cycle. The metric background is of the form 
\beq \label{metric_11d}
   ds^2_{11}=\frac{v^0}{\I\, \tau} \big((d x + \R\, \tau d y')^2 + (\I\, \tau)^2 d y'^2 \big) + ds^2_9 \ ,
\eeq
where $\tau$ is the complex structure modulus of the $T^2$, and $v^0$ describes its volume. 
If the volume $v^0$ of the two-torus is small, one can pick one of the 1-cycles, 
say the A-cycle, to obtain Type IIA string theory. 
T-duality along the B-cycle leads to the corresponding Type IIB set-up, and identifies $\tau$
as the Type IIB dilaton-axion $\tau = C_0 + i e^{-\phi}$.

This construction can be now applied fiber-wise for the elliptically fibered Calabi-Yau 
fourfold $X_4$. Consider M-theory on $X_4$, which leads to a three-dimensional
theory with $\cN=2$ supersymmetry. The reduction and T-duality on the elliptic fiber  
leads\footnote{At least in the absence of metric fluxes; see subsection \ref{Mfluxes}.} to Type IIB string theory on $B_3 \times S^1$, where $B_3$ is the base. Indeed, it is possible to see that the non-triviality of the original elliptic fibration is now encoded by the IIB dilaton-axion profile and thus the Type IIB Einstein frame metric gives simply a direct product geometry.
Let us have a closer look at this set-up from a four-dimensional point of view.
As in \eqref{metricfluct} we label the $S^1$ dimension by $y$ and note that the four-dimensional 
metric is of the form 
\beq \label{4d-metric_red}
   ds^4 = r^{-2} g_{\mu \nu}^{(3)} dx^\mu dx^\nu + r^2 (dy + A_{\mu}^0 dx^\mu)^2\ ,  \qquad \mu,\nu=0,1,2\ ,
\eeq
where $r$ is the circumference of the fourth dimension, $g_{\mu \nu}^{(3)}$ is the 
three-dimensional Einstein frame metric, and $A_\mu^0$ is a three-dimensional vector.
In the following we will identify $(r,A^0)$ with their analogs arising 
in the M-theory reduction on $X_4$.

Let us consider for the moment M-theory on a non-singular
Calabi-Yau fourfold $X_4$ with an elliptic fibration (with 0-section), so that 
there is no non-Abelian gauge 
symmetry in three dimensions. Here and in the rest of the paper we will restrict 
to geometries with full $SU(4)$-holonomy, and demand $h^{2,1}(X_4)=0$. 
On smooth elliptic fibrations there is a natural set of divisors which 
span $H_6(X_4,\bbR)$. Firstly, one has the section of the fibration which is homologous to the 
base $B_3$. Secondly, there is the set of vertical divisors $D_\alpha$ which are 
obtained as $D_\alpha = \pi^{-1}(D_\alpha^{\rm b})$, where $D_\alpha^{\rm b}$ is 
a divisor of $B_3$ and  $\pi$ is the projection to the base $\pi: X_4 \rightarrow B_3$.
For these smooth elliptic fibrations one has $h^{1,1}(B_3)=h^{1,1}(X_4)-1$ such divisors.
Classically Euclidean M5-branes wrapped on such divisors will couple to 
the complex coordinates 
\beq \label{class_T0Talpha}
   T_0 = \frac{1}{6}\int_{B_3} J \wedge J \wedge J + i \int_{B_3} C_6\ ,\qquad 
   T_\alpha = \frac{1}{6}\int_{D_\alpha} J \wedge J \wedge J + i \int_{D_\alpha} C_6 \ ,
\eeq
where $C_6$ is the dual of the M-theory three-form $C_3$.
In order to make $T_0,T_\alpha$ dimensionless 
one would need to multiply the volume terms by $\ell_M^{-6}$.
The three-dimensional kinetic terms of these chiral multiplets are obtained 
from a K\"ahler potential $K^{\rm M}(T+\bar T)$.
To determine $K^{\rm M}$ one analyses the Weyl rescaling to the three-dimensional 
Einstein frame. In a large volume compactification, only the classical volume 
$\cV$ arises as prefactor of the Einstein-Hilbert term. Comparing this with
the $e^{K^{\rm M}}$ prefactor in the scalar potential, one infers~\cite{Haack:1999zv}
\beq \label{class_KM}
   K^{\rm M} = - 3 \log \cV - \log \int \Omega_4 \wedge \bar \Omega_4\ , \quad \qquad \cV = \frac{1}{4!}\int_{X_4} J \wedge J \wedge J \wedge J\ .
\eeq
To evaluate $K^{\rm M}$ as a function of $T+\bar T$ one first expands the K\"ahler form $J$
as
\beq
  J=v^0 \omega_0 + v^\alpha \omega_\alpha\ .
\eeq
where $\omega_0,\omega_\alpha$ are the two-forms Poincar\'e dual to $B_3,D_\alpha$.
Using this expansion one has to solve \eqref{class_T0Talpha} for 
the modes $v^0,v^\alpha$ of $J$ and insert the result into \eqref{class_KM}.
This evaluation is more conveniently performed in a 
dual picture. Simply performing the dimensional reduction, 
one notes that the $v^0,v^\alpha$ 
arise as scalar components of three-dimensional vector multiplets 
$(v^\alpha,A^\alpha)$ and $(v^0,A^0)$, where $(A^0,A^\alpha)$ appear in the expansion of the M-theory three-form 
\beq 
    C_3=A^0 \wedge \omega_0 +A^\alpha \wedge \omega_\alpha\ . 
\eeq

Let us now discuss the map of the M-theory data to the 4d/3d supergravity data of section \ref{3dlandscape} following
\cite{Grimm:2010ks}.
The complex coordinate \eqref{class_T0Talpha} are related to the three-dimensional chiral multiplets 
$T_\Sigma=(T_0,T_\alpha)$ of section \ref{3dlandscape} via a Legendre transform. 
One now notes that $A^0$ is precisely the vector which appears 
in the metric \eqref{4d-metric_red}. Indeed, $C_3$ with two legs along the elliptic fiber becomes the B-field after reduction to Type IIA and the Kaluza-Klein vector after T-duality to Type IIB. Hence, the T-duality 
operation in going from an M-theory compactification to 
a Type IIB compactification is the Legendre transform 
in the direction $T_0$ with respect to the K\"ahler potential $K^{\rm M}$. 
The dual coordinate is given by 
\beq \label{Legendre1}
  R = -2 \partial_{T_0} K^{\rm M} = \frac{v^0}{\cV} \ .
\eeq
One thus identifies the circumference $r$ in \eqref{4d-metric_red} with $R$ 
in the way we have already seen in subsection \ref{3dlandscape}, i.e. 
\beq \label{Rr_id}
    R = r^{-2}\ .
\eeq

One can now evaluate $T_0$ more explicitly by again 
using the fact that $X_4$ is elliptically fibered.
For an elliptic fibration the intersection numbers satisfy 
\beq\label{intersection0}
  \cK_{\alpha \beta \gamma \delta} = D_\alpha \cap D_\beta \cap D_\gamma \cap D_\delta = 0\ ,
\eeq
for the vertical divisors. This allows us to split $T_0$ in a small $R$ expansion as 
\beq \label{T0R}
    \R T_0 = \frac16 \int_{B_3} J\wedge J \wedge J = \frac{1}{R} + p(R) = r^2 + p(r^{-2})\ ,
\eeq
where $p(R)$ is a power series in $R$ with no further poles 
and we have used \eqref{class_T0Talpha}, \eqref{Legendre1} and \eqref{Rr_id}.

In the presence of non-Abelian seven-brane 
stacks the Calabi-Yau manifold itself will become singular due 
to the singularities in the elliptic fibration. 
To nevertheless perform the reduction one can resolve 
these singularities and introduce 
additional two-form classes for the resolution divisors. Such fourfold  
resolutions have recently been studied in refs.~\cite{Blumenhagen:2009yv,Chen:2010ts,Morrison:2011mb,Esole:2011sm,Marsano:2011hv}.
Let us assume for simplicity to have just one such stack and let us call  $\omega_i$ the Poincar\'e duals 
of the exceptional divisors $D_i$, with $i=1\ldots{\rm rk}\,G$, where $G$ is the non-Abelian gauge group 
corresponding to the singularity. Let the 7-brane stack wrap a generic divisor of the base given by 
the Poincar\'e dual in $B_3$ of $\omega_{(\tau)}\equiv \frac12 C^\alpha\omega_\alpha$, $C^\alpha$ being integral coefficients. 
Then, after blow-up, the singular fibration is replaced by the following resolved one:
\be\label{singularFiber}
\hat\omega_{(\tau)}=\omega_{(\tau)}+a^i\omega_i\,,
\ee 
where $a^i$ are the Dynkin numbers characterizing the Dynkin diagram of $G$ and $\omega_{(\tau)}$ 
now corresponds to the fibration of the extended node over the 7-brane stack. Each blow-up divisor is instead a fibration of the corresponding Cartan node over the 7-brane stack.  In addition to \eqref{intersection0}, 
one has the following useful relations for some intersection numbers which involve the blow-up divisors:
\bea\label{intersection_i}
D_0\cap D_i&=&0\,,\nn\\
\cK_{i \alpha \beta \gamma } = D_i\cap D_\alpha \cap D_\beta \cap D_\gamma&=& 0\,,\nn\\
\cK_{ij \alpha \beta} = D_i\cap D_j\cap D_\alpha \cap D_\beta&=& -C^\gamma_{ij}\cK_{\alpha\beta\gamma}\,,
\eea
where $C^\alpha_{ij}\equiv C^\alpha C_{ij}$, the latter being the Cartan matrix of $G$ 
and $\cK_{\alpha\beta\gamma}\equiv \cK_{0\alpha\beta\gamma}$.
Due to the additional harmonic 2-forms, we will now have further 3d vector multiplets 
after compactification, whose bosonic content is given by
\be
(A^i,\xi^i)\,,\qquad\qquad \xi^i\equiv \frac{v^i}{\cV}\,, 
\ee
where, as usual, $A^i$ and $v^i$ come from the expansion along $\omega_i$ of $C_3$ and $J$ respectively.

As opposed to the 4d/3d context, in the M-theory picture it is natural to express all 
the relevant quantities of the effective theory in a formulation where 
the dynamical vector multiplets $(R,A^0)$, $(L^\alpha, A^\alpha)$ 
and $(\xi^i,A^i)$ are kept in the spectrum \cite{Berg:2002es,Grimm:2010ks}, where 
we have set $L^\alpha = v^\alpha/\cV$. 
So instead of working with a K\"ahler potential as in the general form of the 3d action \eqref{class_KM}, one is 
using the Legendre transform 
\beq
   \tilde K^{M}(R,L,\xi) = K^{M} + \tfrac12 (T_\Lambda +\bar T_{\Lambda}) L^{\Lambda} \ , 
\eeq
where $L^{\Lambda} = (R,L^\alpha,\xi^i )$. This expression can be evaluated for $\R T_\lambda = \frac{1}{3!}\int_{D_\Lambda} J^3 $ and $K^M$
as given in \eqref{class_KM}.
Using \eqref{intersection0} and \eqref{intersection_i} one finds   
\bea\label{fullKinetPot}
\tilde K^M=\log R+\log\left[\left(\frac{L^\alpha L^\beta L^\gamma}{6}-\frac{R L^\alpha L^\beta k^\gamma}{4}+\frac{R^2L^\alpha k^\beta k^\gamma}{6}-\frac{R^3k^\alpha k^\beta k^\gamma}{24}\right.\right.\nn\\ 
\left.\left.\;\qquad\qquad-\frac{L^\alpha L^\beta \xi^i\xi^j C^\gamma_{ij}}{4R}\right)\cK_{\alpha\beta\gamma}+\frac{\xi^i\xi^j\xi^k L^\alpha }{6R}\cK_{ijk\alpha}+\frac{\xi^i\xi^j\xi^k\xi^l}{24R}\cK_{ijkl}\right]+4\,.
\eea
In this expression we have introduced the expansion coefficients $k^\alpha$, and used the fact that $\hat X$ is 
a resolved elliptic fibration. More precisely, one has
\beq \label{kalpha_stuff}
   c_1(B_3) = k^\alpha \omega_\alpha|_{B_3}\ , \qquad \omega_0 \wedge \omega_0 = -\hat\pi^* c_1(B_3)\wedge \omega_0
\eeq
with $\hat\pi:\hat X_4\rightarrow B_3$.\footnote{Notice that the expansion coefficients of 
$\hat\pi^*c_1(B_3)$ along $\omega_i$, which in general are non-trivial, do not contribute to $\omega_0^2$ because 
of the first relation in \eqref{intersection_i}.} 
All the volumes here are expressed in units of the 11d Planck length. 
For later purposes, it is useful to write down the precise behavior of these quantities in the small $R$ expansion (F-theory limit). Using the fact that $\cV$ is quartic in the $v$'s with leading term $v^0\cdot(v^\alpha)^3$, and that the $L^\alpha$  remain finite in this limit, it is easy to find the rules \cite{Grimm:2010ks},
\bea\label{rulesFlim}
v^0\sim\epsilon\,,&&R\sim\epsilon^{3/2}\,,\nn\\
v^\alpha\sim\epsilon^{-1/2}\,,&&L^\alpha\sim 
L^\alpha_{\rm b}\equiv\frac{v^\alpha_{\rm b}}{\cV_{\rm b}} 
\,,\nn\\
v^i\sim\epsilon^{3/2}\,,&&\xi^i\sim\epsilon^2\,,
\eea
where the subscript $\rm b$ refers to the corresponding quantities of the base $B_3$.
Notice that the behavior for the $\xi^i$s follows from the fact that 
the $\zeta^i$s of \eqref{4d3dvectors} should be also vanishing in this limit, in 
order not to generate a Poincar\'e-breaking background in four dimensions.

\subsection{M-theory fluxes}\label{Mfluxes}

In M-theory compactifications on Calabi-Yau fourfolds one has to distinguish 
two classes of four-form flux $G_4$. One notes that $H^{4}(X_4,\bbC)$ splits 
into a horizontal and a vertical part, orthogonal to each other with respect to the non-degenerate scalar product given by wedge and integration: 
\beq \label{H4split}
   H^{4}(X_4,\bbC) = H^{4}_{\rm H}(X_4,\bbC)\oplus H^{4}_{\rm V}(X_4,\bbC)\ .
\eeq 
The M-theory fluxes $G_4$ split accordingly.
The elements in $H^{4}_{\rm H}(X_4,\bbC)$ are of varying Hodge-type and the whole 
space $H^{4}_{\rm H}(X_4,\bbC)$ is spanned by the complex structure variations of the holomorphic $(4,0)$-form 
$\Omega$ on $X_4$, such that
\bea
H^{4}_{\rm H}(X_4,\bbC)&=&H^{4,0}\oplus H^{3,1}\oplus H^{2,2}_{\rm H}\oplus H^{1,3}\oplus H^{0,4}\;(X_4,\bbC)\nonumber\\
H^{2,2}_{\rm H}&\subset& H^{2,2}|_{\rm primitive}\;(X_4,\bbC)\;.
\eea
For elliptically fibered Calabi-Yau fourfolds, in this group there are fluxes 
with one and only one leg along the elliptic fiber and as such do not 
break 4d Poincar\'e invariance after the F-theory limit $R\rightarrow0$.  
Such fluxes have been studied in the F-theory context recently in refs.~\cite{Grimm:2009ef,Alim:2009bx,Braun:2011zm}.
In contrast, the elements in $H^{4}_{\rm V}(X_4,\bbC)$ are necessarily of 
Hodge-type $(2,2)$, and can be represented as a wedge product of two elements 
of $H^{1,1}(X_4,\bbC)$. Given the basis $(\omega_0,\omega_\alpha)$ of $H^{1,1}(X_4)$ 
considered in subsection \ref{mod_id} for a smooth elliptic fourfold, we see that 
here there are fluxes with either 2 or no legs along $T^2$ and thus they break 
4d Poincar\'e invariance after the F-theory limit. It is worth to remark here 
that if the elliptic Calabi-Yau has full $SU(4)$-holonomy, its base $B_3$ is not any 
K\"ahler manifold, but it has $h^{0,1}=h^{0,2}=h^{0,3}(B_3)=0$. Hence, the fourth cohomology of $B_3$
splits as $H^4(B_3)=H^{2,2}(B_3)\simeq H^{1,1}\wedge H^{1,1}(B_3)$. This implies that 
all fluxes with 0 or 2 legs along $T^2$ (4d Poincar\'e breaking) lie in the 
second summand of \eqref{H4split}.

Let us further analyze fluxes when $X_4$ is non-singular. 
Since we have given an explicit basis of $H^{1,1}(X_4)$ in subsection \ref{mod_id},
we can classify the parts of $G_4^{\rm V} $ as
\beq\label{G4fluxN}
   G_4^{\rm V} =  N^\alpha\, \omega_0 \wedge \omega_\alpha + \tilde N_\alpha \,
   \tilde \omega^\alpha  \ ,
\eeq 
where $\tilde \omega^\alpha$ are the pull-backs of four-forms on $B_3$ such that 
$\int_{B_3} \omega_\alpha \wedge \tilde \omega^\beta = \delta_\alpha^\beta$.
The coefficients $N^\alpha, \tilde N_{\alpha}$ are appropriately quantized to satisfy \eqref{quantization}. 
Note that the coefficient of the term $\omega_0 \wedge \omega_0$ has been 
absorbed in the $N^\alpha$. 
One also has in cohomology  
\beq\label{KAlphaBetaGamma}
   \omega_\alpha \wedge \omega_\beta  = \cK_{\alpha \beta \gamma}  \tilde \omega^\gamma \ , \qquad \quad \cK_{\alpha \beta \gamma} = \int_{B_3} \omega_\alpha \wedge \omega_\beta \wedge \omega_\gamma\ , 
\eeq
such that the $\tilde N_\alpha$ parameterize independent degrees of 
freedom where the relations among the $\omega_\alpha \wedge \omega_\beta$ 
have been modded out. 

The integers $N^\alpha$ correspond to the Chern class of an $S^1$-fibration on $B_3$ 
integrated over the 2-cycles PD$_{B_3}(\tilde\omega^\alpha)$ in Type IIB. Note that this is consistent 
with performing the Type IIA limit of M-theory and performing a T-duality around 
the second circle in the elliptic fiber. In the IIA limit $N^\alpha$ labels $H_3$ NS-NS 
fluxes which, as is well known, are translated to metric fluxes under T-duality along 
one of their legs. The $S^1$ fiber is the T-dual of the $B$-cycle, which we call $B_T$ in the following.
Such fluxes make the mere existence of the 4d effective theory questionable, as this circle constitutes the third spatial direction of the 4d space-time.
We will later demand that these components of $G_4$ vanish.

In contrast, the $\tilde N_\alpha$ correspond to the flux quanta of the axions obtained by expanding 
$C_4$ in four-forms of $B_3$ in the Type IIB picture as  
\beq\label{Ntildealpha}  
  \tilde N_\alpha  = \int_{B_T} d\, \I \, t_\alpha \ , \qquad C_4 = \I\, t_\alpha \, \tilde \omega^\alpha |_{B_3}\,,
\eeq
where $\I t_\alpha$ are real 4d scalars.  Such fluxes do not induce any non-trivial fibration of $B_T$ on $B_3$ and are S-duality invariant. 
Note that the fields $t_\alpha$ are related to the fields $T_\alpha$ of section \ref{from4to3} and \ref{mod_id}, whose imaginary parts are 3d dual to the $A^\alpha$, 
by the relation 
\be
T_\alpha=\frac{t_\alpha}{2}\,,
\ee
as can be checked by comparing the $\cN=1$ gauge coupling functions \cite{Grimm:2011tb}. 
The $1/2$ factor is compatible with the gauge transformation of $C_6$ in 11d supergravity
\be\label{C6gaugetransf}
C_6\longrightarrow C_6+\cfrac12\Lambda_2\wedge G_4\,,
\ee
where $\Lambda_2$ defines the gauge transformation for $C_3$, i.e.~$C_3\to C_3+d\Lambda_2$. Eq. \eqref{C6gaugetransf} induces a $1/2$ in the gaugings of the shift symmetry of $\I T_\alpha$, because these gaugings arise from the term 
$\frac{1}{2} \,C_3\wedge G_4$ integrated over the vertical divisors of $X_4$. Therefore, one has
\be
M_\alpha=\frac{\tilde N_\alpha}{2}\,,
\ee 
where $M_\alpha$ define the gaugings we considered in subsection \ref{3dlandscape} in the context of the 4d/3d reduction.\footnote{When the Calabi-Yau fourfold is singular there is an additional contribution to the 4d fluxes $M_\alpha$, as it will be shown in subsection \ref{Minterpret}.}

In the presence of a non-Abelian singularity one has additional vertical fluxes. The complete expansion of $G_4^V$ reads
\beq\label{G47branes}
   G_4^{V} = N^\alpha\, \omega_0 \wedge \omega_\alpha + \tilde N_\alpha \tilde \omega^\alpha+f^{i\alpha}\, \omega_{i} \wedge \omega_{\alpha} + \tilde f^{ij}\, \omega_{i} \wedge \omega_{j}\ .  
\eeq
The two new terms will be crucial for the M-theory interpretation of the 4d/3d reduction with vectors performed in subsection \ref{4dVectors}. Following the duality recalled in subsection \ref{mod_id} to link M-theory on $X_4$ with Type IIB on $S^1 \times B_3$ 
one expects that the $f^{i\alpha}$ map to the periods of a seven-brane two-form flux $F_2^i$ 
along the $i$-th Cartan direction of the non-Abelian gauge group $G$. 
They exist in the four-dimensional gauge theory in the form of the gaugings $X_{i\alpha}$ 
introduced in subsection \ref{4dVectors} and are not special to three-dimensions. 
As we will show below, this matching will be modified by the fluxes $\tilde f^{ij}$.

Let us close this subsection by commenting on the quantization of the fluxes in \eqref{G47branes}. 
Notice that  $c_2(X_4)$ lies in the second summand 
of \eqref{H4split}, and the $G_4$ fluxes in this part of the cohomology 
have to satisfy a shifted quantization condition \cite{Witten:1996md}. 
Vertical fluxes are in fact quantized in the following way
\beq \label{quantization}
    G^{\rm V}_4 + \frac{c_{2}(X_4)}{2} \ \in \ H^{4}_{\rm V}(X_4,\bbC) \cap H^4(X_4,\bbZ)\ .
\eeq 
This condition has been 
recently studied explicitly for elliptic, possibly singular, 
fourfolds \cite{Collinucci:2010gz}. In contrast, horizontal fluxes are never affected by non-trivial 
quantization conditions.

\subsection{The D-term potential}\label{Minterpret}
 
We are now ready to present an M-theory realization of the terms of the 3d effective 
action derived in subsections \ref{3dlandscape} and \ref{4dVectors} by flux compactification of a 4d, shift symmetric, $\cN=1$ theory. 
  
While the fluxes $G_4^{\rm H}$ in $H^{4}_{\rm H}(X_4,\bbC) \cap H^4(X_4,\bbZ)$ appear in the Gukov-Vafa-Witten superpotential \cite{Gukov:1999ya}
\beq
   W_{\rm GVW} = \int_{X_4}\Omega_4 \wedge G_4^{\rm H}\,,
\eeq  
the vertical fluxes appear in three dimensions in the function \cite{Haack:1999zv}
\beq\label{DPotentialM}
   \cT = \frac{1}{8\cV^2} \int_{X_4} J \wedge J \wedge G_4^{\rm V}\ .
\eeq
They determine the 3d scalar potential as in eq. \eqref{3dPotential}.

Let us focus on the K\"ahler moduli of the M-theory compactification, which correspond to the fields $T_\Sigma$ of section \ref{from4to3}. Using  \eqref{G47branes}, the D-term potential \eqref{DPotentialM} reads:
\be\label{TPotentialTildeN}
\cT = \frac{1}{8} \int_{\hat X_4} \left(R\omega_0+L^\alpha\omega_\alpha+\xi^i\omega_i\right)^{2}\wedge\left(N^\beta\, \omega_0 \wedge \omega_\beta+\tilde N_\gamma \tilde \omega^\gamma+f^{j\delta}\omega_j\wedge\omega_\delta +\tilde f^{kl}\, \omega_{k} \wedge \omega_{l}\right) 
\ee
where the integral is performed on the blown-up Calabi-Yau fourfold. After using \eqref{intersection0} and \eqref{intersection_i} to simplify the expression \eqref{TPotentialTildeN}, one can easily read off the embedding tensor 
$\Theta_{\Lambda\Sigma}$ in front of the term $L^\Lambda L^\Sigma$, i.e.~we have
\be\label{G4Gaugings}
\Theta_{\Lambda\Sigma}=-\cfrac14\int_{\hat X_4} G_4\wedge\omega_\Sigma \wedge \omega_\Lambda\,.
\ee
It is then clear that in general $\Theta_{0\alpha}$ and $\Theta_{i\alpha}$ 
will not be the only non-zero components. This is due to the fact, that the flux content and the geometric 
structure of the full M-theory compactification are much richer than the ones we had in 
the 4d theory discussed in section \ref{from4to3}. Therefore, in order that the present M-theory 
compactification leads to a three-dimensional theory 
which can also be obtained by a 4d/3d flux compactification, 
as performed in subsection \ref{4dVectors}, we have to impose the following two restrictions:
\begin{enumerate}
\item We constrain some of the fluxes in order to ensure the vanishing of the components of the embedding tensor which have 
no analog in the reduction from 4d to 3d presented in section \ref{4dVectors}. Namely, we impose the conditions (see eq. \eqref{ThetaX0})
\be
\Theta_{\alpha\beta}=\Theta_{ij}=\Theta_{00}=\Theta_{0i}=0\, .
\ee
\item We discard the terms in the kinetic potential \eqref{fullKinetPot} 
     which  lead in the 3d scalar potential to terms $\cO(\epsilon^{x>4})$ in the F-theory limit \eqref{rulesFlim}.
\end{enumerate}
The second restriction comes from the fact that, after reduction of the 4d theory 
with vectors and the Weyl rescaling $g^{(3)}\rightarrow Rg^{(3)}$, the most 
suppressed term we get in the large $S^1$ expansion of the 3d scalar 
potential looks schematically like $R^2\zeta^2\sim\cO(\epsilon^4)$, where 
the first factor is due to the Weyl rescaling. From the direct computation of the 
scalar potential given in appendix \ref{Dualization4dVect}, one easily infers that, 
to satisfy such a condition, we need to keep those terms in \eqref{fullKinetPot} 
which are at most of order $\cO(\epsilon^{5/2})$. The term linear in the first Chern 
class, which by the rules \eqref{rulesFlim} still survives, can be easily eliminated 
by the following field redefinition\footnote{Remarkably, such field redefinition 
eliminates also the term of order three in the first Chern class, which in our contest is irrelevant.}
\bea\label{FieldRedef}
L^\alpha\longrightarrow L^\alpha+\frac{k^\alpha}{2}R\;,&\qquad\qquad&
A^\alpha\longrightarrow A^\alpha+\frac{k^\alpha}{2} A^0\,,
\eea
with all the other fields unchanged. We can then regard the new fields as the right 3d K\"ahler coordinates in terms of which the 4d/3d reduction goes exactly as described in subsection \ref{3dlandscape}. Expanding \eqref{fullKinetPot} according to \eqref{rulesFlim} and applying restriction 2 above, the kinetic potential, in the new basis of fields, reads
\be\label{KineticPotentialVect}
\tilde K=\log R-\frac{1}{R}{\rm Re}\,\tau_{ij}\xi^i\xi^j+\log\big(\tfrac16 L^\alpha_{\rm b} L^\beta_{\rm b} L^\gamma_{\rm b} \cK_{\alpha\beta\gamma}\big)+4\,,
\ee
where we have used that the leading order term for $\tau$ is given by
\be\label{RelevantKinet}
\tau\sim\frac{L^\alpha_{\rm b} L^\beta_{\rm b} C^\gamma}{4 \cV^{-2}_{\rm b}}\cK_{\alpha\beta\gamma}\quad\qquad{\rm with}\quad\qquad\cV_{\rm b}^{-2}=\tfrac16 L^\alpha_{\rm b} L^\beta_{\rm b} L^\gamma_{\rm b} cK_{\alpha\beta\gamma}\,.
\ee 
This is compatible with the fact that $\tau=\frac12 C^\alpha T_\alpha$. 
It is easy to verify that the kinetic potential \eqref{RelevantKinet} leads 
via Legendre transformation exactly to the complex coordinates defined 
in \eqref{TiFields} and \eqref{T0Field}, as well as to the K\"ahler 
potential \eqref{KaehlerPotVect}.\footnote{One also neglects derivatives of $\tau$ 
with respect to the $L$'s, because they lead to higher order corrections in $\epsilon$.}
Further details on this theory, including its dualization to the vector multiplet formalism, are given in appendix \ref{Dualization4dVect}.

Let us now come back to restriction 1 above, and deduce the right constraints on fluxes. After the field redefinition \eqref{FieldRedef}, the expressions of the various components of the embedding tensor in terms of the flux quanta of $G_4$ read
\bea \label{all_the_thetas}
\Theta_{00}&=&-\tfrac{1}{16}\big(N^\alpha-\tilde f^{ij}C^\alpha_{ij}\big)k^\beta k^\gamma\cK_{\alpha\beta\gamma}\,,\nn\\
\Theta_{0\alpha}&=&-\tfrac14 \tilde N_\alpha+\cfrac18\big(N^\beta+\tilde f^{ij}C^\beta_{ij}\big)k^\gamma\cK_{\alpha\beta\gamma}\,,\nn\\
\Theta_{\alpha i}&=&-\tfrac14 \big(\tilde f^{jk}\cK_{\alpha ijk} -f^{j\beta}C^\gamma_{ij} \cK_{\alpha\beta\gamma}\big)\,,\nn\\
\Theta_{0i}&=&\tfrac{1}{2} k^\alpha\Theta_{\alpha i}\,,\nn\\
\Theta_{\alpha\beta}&=&-\tfrac14\big(N^\alpha-\tilde f^{ij}C^\alpha_{ij}\big)\cK_{\alpha\beta\gamma}\,,\nn\\
\Theta_{ij}&=&-\tfrac14\big(f^{k\alpha}\cK_{\alpha kij}+\tilde f^{kl}\cK_{ijkl}-\tilde N_\alpha C^\alpha_{ij}\big)\,,
\eea
where $\cK_{\alpha ijk}$ and $\cK_{ijkl}$ are new intersection numbers whose explicit forms in terms of invariants of the group $G$ are not relevant for our purposes. Imposing restriction 1 amounts to two further constraints on the fluxes given by
\bea
N^\alpha-\tilde f^{ij}C^\alpha_{ij}=0\label{ConstraintNalpha}\,,\\
\tilde f^{kl}\cK_{ijkl}=\tilde N_\alpha C^\alpha_{ij}-f^{k\alpha}\cK_{\alpha kij}\label{Constraintfij}\,.
\eea
The conditions \eqref{ConstraintNalpha} and \eqref{Constraintfij} respectively specify the 
fluxes $N^\alpha$ (explicitly) and $\tilde f^{ij}$ (implicitly) in 
terms of $\tilde N_\alpha$ and $f^{k\alpha}$. However, the latter remain completely unfixed. 
In this way we achieved $\Theta_{\alpha\beta}=\Theta_{ij}=\Theta_{00}=0$. 

There is one further constraint which we still need to impose, $\Theta_{0i}=0$. However, 
as already given in \eqref{all_the_thetas}, this component obeys  $\Theta_{0i}=\frac{1}{2} k^\alpha \Theta_{\alpha i}$.
The latter identity can be inferred by realizing that the $k^\alpha$ are related to the first Chern class of $B_3$
and appear in the geometrical constraint \eqref{kalpha_stuff}. The vanishing constraint then reads 
\beq \label{ktheta=0}
   k^\alpha \Theta_{\alpha i} = 0\ .
\eeq
It is interesting to relate this condition to a 4d constraint. In order do that 
it is necessary to read off the $k^\alpha$ in a purely 4d context. It was argued 
in \cite{GrimmTaylor} that this can be done using the higher curvature terms \eqref{def-cLR2}. In fact, 
the coefficients $k^\alpha$ appearing in the expansion of the 4d scalar $\sigma$, eq. \eqref{simple_sigma}, 
coincide with the ones of the expansion of the first Chern class of the M-theory fourfold. 
This implies that the constraint \eqref{ktheta=0} has to be identified with the second condition in 
\eqref{Assumption} by using $\Theta_{\alpha i} = \frac{i}{2} X_{i\alpha}$ as in \eqref{ThetaX0}.

The non-vanishing components $\Theta_{0\alpha}$ and $\Theta_{i\alpha}$ are given in terms of the 
unconstrained fluxes $\tilde N_\alpha,\, f^{i\alpha}$ together with $\tilde f^{ij}$ solving eq. \eqref{Constraintfij}. 
To summarize, except for the conditions \eqref{Assumption}, which already exist in 4 dimensions,  
$\Theta_{i\alpha}$ is completely arbitrary, exactly as it is in the 4d theory for $X_{i\alpha}$. 
Analogously, recalling equations \eqref{kthetaM} and \eqref{CQ=0M} for the $M_\alpha$ in the 4d/3d 
reduction, $\Theta_{0\alpha}$ must be such that $C^\alpha\Theta_{0\alpha}=k^\alpha\Theta_{0\alpha}=0$. 

Moreover, let us stress that, when non-Abelian singularities are present, the condition of absence of ``metric'' fluxes, which in the smooth situation is simply $N^\alpha=0$, becomes eq. \eqref{ConstraintNalpha}. Similarly, the 7-brane gauge flux, which is related to $\Theta_{\alpha i}$ gets contributions also from $\tilde f^{ij}$ fluxes and not only from $f^{i\alpha}$. As a final comment, notice that in the case $h^{1,1}(B_3)=1$, the 4d conditions \eqref{Assumption} are equivalent to absence of 4d gaugings.

\subsection{M5-brane instantons, Taub-NUTs and Fluxes} \label{Taub-NUT_M}

In this final subsection we provide the M-theory realization of the gravitational and gauge 
instanton corrections discussed in subsections \ref{StabRadion} and \ref{FluxSuperpotential}. M-theory unifies all 
these corrections as coming from a single source: M5-branes wrapped on divisors of the blown-up 
Calabi-Yau fourfold $\hat{X}_4$.

Let us first concentrate on the simplest situation of a Taub-NUT instanton without 
fluxes correcting the 3d superpotential as in \eqref{Taub-NUT_super}. 
In ref.~\cite{Witten:1996bn} it has been studied when M5-brane instantons have 
the right number of zero-modes to contribute to the superpotential $W$ of the 
3d effective theory. For M-theory on an elliptically fibered Calabi-Yau fourfold with 0-section, and 
in the absence of $G_4$-fluxes, there 
is always an M5-brane wrapped on the base $B_3$ which contributes to 
the superpotential. Using the fact that that $h^{i,0}(X_4)=0,i=1,2,3$ 
one concludes that $h^{i,0}(B_3)=0$ for the base $B_3$ of an elliptic fibration. In particular, 
this implies that the holomorphic Euler characteristic satisfies $\chi_h(B_3) = \sum_i (-1)^i h^{i,0}=1$. 
Accordingly, the M5-brane with action $T_0$ given in \eqref{class_T0Talpha}, \eqref{T0R} contributes a
superpotential of the form $W = \cA e^{-2\pi T_0}$. 
Note that upon reduction and T-duality this M5-brane instanton turns into a Taub-NUT gravitational instanton. 
Indeed, reducing M-theory to Type IIA on the $A$-cycle the M5-brane reduces to an NS5-brane.
After T-duality along the $B$-cycle the NS5-brane is dualized into a Taub-NUT geometry. More precisely, 
the 10d Type IIB dual to the M5-brane instanton background is ${\rm TN}\times B_3$. This duality explains the 
precise match of the 3d M-theory superpotential and the superpotential \eqref{Taub-NUT_super}
discussed in subsection \ref{StabRadion}.

It is important to remark that in the decompactification limit 
induced by shrinking the torus fiber of $X_4$, the F-theory limit $R \rightarrow 0$, 
the action \eqref{T0R} of the M5-brane on $B_3$ becomes infinite and the corresponding correction 
to the superpotential is absent. The dual 4d analog is the fact that the Taub-NUT action 
is infinite in the limit  $r\rightarrow\infty$, in which the metric \eqref{MetricTaubNUT} becomes the flat metric on $\mathbb{R}^4$. 

Let us now include a non-trivial $G_4$-flux into the discussion of the instanton corrections. 
This will be the M-theory realization of the setup described in subsection \ref{FluxSuperpotential},
and provide a microscopic point of view of the compatibility between fluxes and the gravitational instantons. 
In section \ref{from4to3} we have seen that circle fluxes induce 
non-trivial gaugings for the shift symmetries of the scalars $T_0,T_\alpha$. 
This in turn forbids the appearance of a superpotential of the form $\cA e^{-2 \pi T_0}$ with constant $\cA$ since 
this would break the gauge symmetries. 
Therefore a natural question to ask is, to what the lack of gauge invariance is corresponding to for the M5-brane sources. 
On general grounds, the background fluxes responsible for this phenomenon should 
restrict non-trivially to the worldvolume of the brane sources.  For instance, in a string background, D-branes with a 
non-trivial H-flux on their worldvolume can suffer from a Freed-Witten anomaly \cite{Freed:1999vc}, and thus 
cannot modify the effective action as instanton corrections.

In our case, we need the generalization of the Freed-Witten anomaly to the M5-branes of M-theory.   
This was conjectured by Witten \cite{Witten:1999vg}, and takes an analogous formulation to the string case. 
An M5-brane is non-anomalous if and only if the $G_4$ flux, when restricted to the M5-brane worldvolume, 
satisfies the following condition:
\be
G_4|_{\rm M5}=\theta_4({\rm M5})\,,
\ee
where $\theta_4({\rm M5})\in {\rm Tor}\,H^4({\rm M5},\mathbb{Z})$ is a 2-torsion class of the M5-brane. The four-form 
$\theta_4$ generalizes the third Stiefel-Whitney class $W_3$ of the D-brane case. 
We expect that $\theta_4$ vanishes for complex manifolds, as $W_3$ does. 
Therefore, the M5 brane wrapping the base  $B_3$  does suffer from this anomaly,
because of the non-vanishing $G_4$ flux \eqref{G47branes} on its worldvolume.
In particular, the $G_4$ fluxes dual to the circle fluxes pull back non-trivially 
to $B_3$.

However, in subsection \ref{FluxSuperpotential}, we have seen a way to cure the 
lack of gauge invariance of the term $\cA e^{-2 \pi T_0}$, by allowing the prefactor $\cA$ 
to depend on the moduli $T_i$. As the form of superpotential \eqref{FieldDependB} suggests
the modified correction should no longer be simply generated by an M5-brane 
instanton wrapping the $B_3$, but by an M5-brane wrapping a more complicated 
divisor $\cD$ of the resolved fourfold $\hat X_4$. Cohomologically the 
Poincar\'e dual two-form to $\cD$ is 
\be
{\rm PD}_{\hat X_4}(\cD)=\omega_0-n^i\omega_i\,.
\ee
The M5-branes wrapping $B_3$ and the blow-up divisors $D_i$ dual to $\omega_i$ are separately anomalous, due to the 
fluxes $\Theta_{0\alpha}$ and $\Theta_{i\alpha}$ respectively. In contrast the M5-brane 
wrapping $\cD$ is anomaly free if the constraint 
\beq\label{0G4}
  G_4|_{\cD} =0:\qquad \Theta_{0\alpha}-n^i\Theta_{i\alpha}=0\, ,
\eeq
is satisfied. Using the duality to the 4d setup of section \ref{from4to3}, we recognize \eqref{0G4} to be the constraint 
\eqref{GaugeInvW} ensuring a gauge-invariant superpotential.
Concerning the first two terms appearing in the 3d superpotential \eqref{FieldDependB}, they are gauge invariant by themselves, thanks to the conditions $C^\alpha\Theta_{i\alpha}=k^\alpha\Theta_{i\alpha}=C^\alpha\Theta_{0\alpha}=k^\alpha\Theta_{0\alpha}=0$, which are the analogs of eqs. \eqref{kthetaM}, \eqref{CQ=0M} and \eqref{Assumption}.  Hence, in the present language, the corresponding M5-brane sources are anomaly free.
Let us finally remark that in case there exists only one K\"ahler modulus, namely  $h^{1,1}(B_3)=1$, in order the M5-brane 
on $B_3$ to be non-anomalous, or equivalently the Taub-NUT instanton contribution to be gauge invariant, 
one has to turn off also the flux for Im$T$ due to eq. \eqref{GaugeInvW}, ending up with a fluxless 
4d/3d reduction.

\begin{figure}[h!] 
  \centering
      \includegraphics[width=4cm]{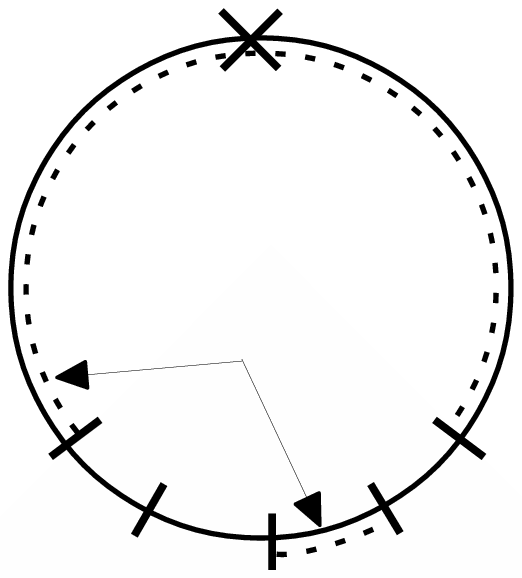}
  \caption{Brane configuration along the Type IIA compactification circle.} \label{branesoncircle}
  \begin{picture}(0,0)
 \put(-055,70){D6$_1$}
 \put(40,70){D6$_N$}
 \put(0,145){NS5}
  \put(5,90){D4s}
\end{picture}
\end{figure}

It is illuminating to also give the Type IIA interpretation of the M5-brane wrapping $\cD$. 
We do that by reducing M-theory along a small $A$-cycle of the torus fiber. 
The result is a system of NS5-D6-D4-branes wrapping various cycles in $B_3$. 
The $B$-cycle of the torus is still fibered over $B_3$, and the various branes 
are points or segments on this circle as depicted in figure \ref{branesoncircle}. More precisely, 
the D6-branes and NS5-branes are points in the $B$-circle, with the NS5-brane sitting 
on the marked point of the circle obtained by the 0-section of the M-theory elliptic fibration.
The support of any D4-brane is a fibration of a line segment in the $B$-cycle over the divisor 
$S = \cD \cap B_3$. These line segments either connect 
two neighboring D6-branes or a D6-brane with the NS5-brane.
The singularity  of the intersection between the D4-brane and the NS5-brane 
is only an artifact of the weak coupling limit \cite{Witten:1997sc}, as it is smoothened 
by growing the $A$-cycle in M-theory, where the two branes both become M5-branes, 
intersecting on a smooth holomorphic surface (diffeomorphic to $S$). 
Upon T-duality along the $B$-cycle, the D6-branes map to D7-branes on ${\rm TN}\times S$ 
and the D4-branes to D3-branes on $S$.

Let us now use this brane setting to reproduce the configuration of gauge instantons 
on Taub-NUT presented in subsection \ref{FluxSuperpotential}. We consider $N$ D6-branes 
on $\bbR^3\times S$. If they are coincident on the $B$-cycle they give rise to an 
$U(N)=SU(N)\times U(1)$ gauge theory on their worldvolume. In contrast, if their positions 
along the circle are generic, the gauge theory is in its Coulomb phase and the group 
is broken to the Cartan torus $U(1)^N$. Notice that this is a different breaking with 
respect to the one seen in subsection~\ref{FluxSuperpotential}, which is due to gaugings, 
arising due to brane fluxes. Here, instead, the breaking arises from the resolution of the 
M-theory singular fiber or, in the dual Type IIB, from a non-trivial Wilson line 
background of the instanton connection. 

Let us first focus on the non-Abelian phase of the gauge theory. T-duality along 
the $B$-cycle dualizes the NS5-brane into a Taub-NUT space with a gauge instanton bundle $E$ on it. 
Adapting the construction of \cite{Cherkis:2008ip,Witten:2009xu} to this situation, we can 
specify the topology of this bundle in terms of the brane construction just described.  
\begin{itemize}
\item
The first Chern class of the instanton bundle vanishes:
\be\label{FirstCh}
{\rm ch}_1(E)=0\,.
\ee
This is due to the fact that, when we have only one NS5-brane localized on a circle, there is no invariant information 
behind its relative position (or its linking number) with respect to a localized D6-brane, as we can move the D6-brane 
all along the $S^1$ always avoiding Hanany-Witten transitions. This is clearly consistent with the fact that $H^2({\rm TN},\mathbb{Z})=0$.
\item
The second 
Chern character of the instanton bundle is given by 
the number of ``complete'' D4-branes, namely those which wrap 
the entire $B$-cycle (i.e.~from the NS5 back to itself). They 
lift to M5 instantons wrapping the entire singular fiber. In 
the singular limit for the blown-up fourfold $\hat X_4$, when the full non-Abelian gauge theory  on the stack of D7-branes is restored, the class $\omega_{(\tau)}$ represents 
the full singular fiber over $S$. Consequently, $e^{-2 \pi \bar n\tau}$ is 
the contribution of $\bar n$ complete D4-branes and hence we have equation \eqref{nBar} as  the second Chern character of $E$.
\end{itemize}

Let us now introduce a Wilson line background for the instanton connection $\cA$. As already stated, 
this will generically break $U(N)$ to $U(1)^N$. 
Therefore, T-duality will now yield  a system of Abelian instantons 
on Taub-NUT. Their properties can be summarized as follows.
\begin{itemize}
\item
The monodromy of the instanton connection $\cA$ along the circle $B_T$ at infinity of TN is
\be\label{instantonHolonomy}
e^{2\pi i\cA}={\rm diag}\left(e^{2\pi i\langle s^1\rangle}, \ldots, e^{2\pi i\langle s^N\rangle}\right)\,,
\ee
where $\langle s^I\rangle$ is the position of the $I$-th D6-brane along the asymptotic $B$-cycle and 
has been defined in equation \eqref{cAexpand}. These $N$ positions are related to the $N-1$ 
$\langle \zeta^i \rangle$ used in section \ref{FluxSuperpotential} via $\langle \zeta^i \rangle = \langle s^i\rangle - \langle s^{i+1}\rangle $.

\item
The second Chern character of the total instanton configuration gets now further contributions. 
In the Coulomb phase, the contribution $\bar n$ replacing eq.~\eqref{nBar} is only due to the 
Abelian field strength corresponding to the central $U(1)$ inside $U(N)$. 
More precisely, it coincides with the number of D4-branes wrapping the segment between the two outer D6-branes, i.e.~D6$_1$ and D6$_N$, 
containing the NS5-brane (see figure \ref{branesoncircle}).
These lift in M-theory to M5-branes wrapping the extended node of the resolved fiber. In addition,
 we have all the terms in the action \eqref{KineticGaugeInstanton}, which provide 
continuous contributions depending on the relative positions $\langle \zeta^i \rangle$ of the various D6-branes along the $B$-cycle. In total we have
\be
{\rm ch}_2(E) = \bar{n}+ n^i n^j C_{ij} + 2 n^i C_{ij} \langle\zeta^j\rangle + C_{ij} \langle\zeta^i\rangle \langle\zeta^j\rangle\,,
\ee
where, for simplicity, we have set to zero the position along the circle of the center of mass of the D6-brane stack, i.e.~$\sum_i\langle s^i\rangle$. Here $n^i$ is interpreted as the number of D4-branes stretching along the $i$-th segment joining two consecutive D6-branes. 
These lift in M-theory to M5-branes wrapping the $i$-th Cartan node of the Dynkin diagram of $SU(N)$.
\end{itemize}

Let us conclude this section with two remarks. One should not forget 
that the $B$-cycle is fibered over $B_3$, while its T-dual $B_T$ is not. Thus the 
whole discussion above should be applied fiberwise. However, topological 
quantities, like linking and winding numbers, do not change as we slide 
the $S^1$ along the base. This is confirmed by the fact that they get 
mapped to Chern characters of the instanton bundle by T-duality. In contrast, 
continuous information like brane locations along the circle 
will depend on the internal point and this is reflected after T-duality 
in a non-trivial dependence of the instanton configuration $\cA$ on the 
internal coordinates. 

As a second remark, we stress that in the more complicated M5-brane setups we have analyzed mainly 
the aspect of anomalies. However, another necessary condition for 
an M5-brane instanton to contribute to the superpotential is that the M5-brane instantons 
have the correct number of fermionic zero modes \cite{Witten:1996bn,Blumenhagen:2009qh}. It would be interesting 
to address this question for the M5-brane configuration on~$\cD$. M5-brane instantons, relevant for F-theory compactifications to four dimensions, 
and their zero mode structure have recently been studied in \cite{Blumenhagen:2010ja,Donagi:2010pd,Cvetic:2010ky,Marsano:2011nn,Bianchi:2011qh}.


\section{Conclusions}

In the first part of the paper we performed a circle compactification with 
fluxes starting with a class of 4d, $\cN=1$ supergravity theories, and 
analyzed in detail the resulting 3d, $\cN=2$ effective actions 
obtained by Kaluza-Klein reduction. The original 4d theories
typically arise from flux compactifications of Type IIB string theory 
with 7-branes, or their F-theory counterparts. 
The non-Abelian gauge symmetry can be broken already in 4d 
to the Cartan torus by gauging the shift symmetries of the scalars. 
The 4d/3d fluxed reduction includes non-trivial vacuum expectation values for the 
field strengths of the shift symmetric 4d scalars. 
After reduction the 4d $S^1$-diffeomorphisms become 3d 
gauge symmetries with the graviphoton as gauge boson, and we have shown 
that the circle fluxes yield additional 3d gaugings of a subset of 3d scalars
under this extra gauge symmetry.

The three dimensional action can be written in a compact form upon dualizing 
all degrees of freedom to complex scalars, while keeping non-dynamical 
vectors only in order to perform the gaugings.
According to the general structure of an 3d, $\cN=2$ gauged supergravity theory, 
the action is defined by three functions of the scalars: The K\"ahler potential $K$, 
the D-term potential $\cT$, and the holomorphic superpotential $W$. 
Due to the gaugings, a non-trivial constant embedding tensor appears, 
inducing Chern-Simons terms for the vector fields. We have evaluated the 
3d characteristic data for our $S^1$ compactification as a function 
of the 4d characteristic data and the circle fluxes. 

Elaborating on the 4d to 3d reduction, we have then 
discussed certain non-perturbative corrections to the 3d 
effective theory. More precisely, 
we have argued that 4d gravitational instantons 
with the geometry of a Taub-NUT space provide a non-trivial contribution to 
the 3d effective superpotential. Such a correction 
depends exponentially on the radion field and is of the form~$\cA e^{-2 \pi T_0}$. 
In the absence of gaugings, such a term allows us to implement 
the usual KKLT procedure in order to fix the value of the radion at an AdS$_3$ minimum. We have
discussed various aspects of such a stabilization scheme. In particular, the backreaction 
of a stepwise moduli stabilization will deform the geometry of the gravitational instanton
to Taub-NUT-AdS or Taub-NUT-dS. 
In a second step one might hope that this vacuum can be uplifted to a metastable 3d de Sitter vacuum.
If the existence of such a dS$_3$ state can be established, as it would be the case if one believes in the 
KKLT up-lift, one can study the tunneling from an effectively 3d theory to an 4d cosmological vacuum.
It would be interesting to explore this possibility in more detail. The 4d/3d perspective will 
shed some new light on the backreaction of stepwise moduli stabilization and the generation 
of de Sitter vacua influences the compactification geometry.
 
Focusing on the inclusion of fluxes, one can study how the gravitational instanton 
corrections contribute in the presence of 3d gaugings induced by circle fluxes. 
We have shown that these gaugings forbid 
a contribution $\cA e^{-2 \pi T_0}$ to the 3d superpotential if $\cA$ is field independent due to its lack of 
gauge invariance. This conclusion can be avoided by considering a field-dependent prefactor 
containing the terms $\cA\propto e^{n^iT_i}$. For suitably chosen integers $n^i$s gauge invariance is restored. 
This setup corresponds to a system of Abelian gauge instanton bundles on top of the Taub-NUT space. Within 
such a configuration we have discussed the structure of the superpotential and commented on the 
origin of contributions arising from higher 4d curvature terms, and the central $U(1)$ in a non-Abelian 
$U(N)$ gauge theory.

In the second part of the paper, we have provided an M-theory realization of the 
4d/3d flux compactifications with gravitational instanton corrections. 
We studied the compactification of M-theory on a singular, elliptically fibered Calabi-Yau fourfold 
with $G_4$ fluxes. The resulting 3d effective theory can be matched with the gauged supergravity 
obtained by the 4d/3d flux compactification if appropriate restrictions are imposed on both 
setups. Firstly, the 4d theory used in the circle reduction has to arise from a Type IIB or F-theory 
compactification, such that the K\"ahler potentials and 4d gaugings can be matched with the M-theory gaugings.
Secondly, also the M-theory reduction has to be restricted, since not all $G_4$ fluxes have a 4d/3d interpretation.
We have found the precise constraints on the $G_4$ flux in order to match the 3d gaugings derived in the context 
of the 4d/3d fluxed reduction. Moreover, a specification of the scaling behavior of the various fields in 
the F-theory limit allowed us to consistently match the M-theory K\"ahler potential, with 
the K\"ahler potential arising in a 4d/3d reduction.

In the last part of the paper we have also given an M-theory 
dual of the Taub-NUT gravitational instanton corrections. 
By reducing M-theory to Type IIA at weak coupling 
and performing one T-duality, the M5-brane wrapped on the base $B_3$ of the 
elliptically fibered fourfold was mapped to the Taub-NUT geometry. 
In the absence of gaugings such an M5-brane universally corrects the 
3d superpotential. The precise match of the 3d gaugings and $G_4$ fluxes 
allowed us then to provide a microscopic explanation for 
the lack of gauge invariance of the effective instanton contribution in 
the presence of a flux background. We have shown that the fluxes inducing 
genuine 3d gaugings restrict non-trivially to the worldvolume of the M5-brane 
instanton on $B_3$. This induces an anomaly on the M5-brane which 
is analogous to the Freed-Witten anomaly for D-branes. 
This anomaly can be avoided by replacing the M5-brane on $B_3$ with 
an M5-brane which has a worldvolume $\cD$ with a more involved geometry. 
Topologically $\cD$ is specified by a number of integers $n^i$ which 
parameterize the wrappings of $\cD$ on the resolution divisors of a non-Abelian 
singularity of the fourfold $X_4$. For appropriate choices of $n^i$, 
a given $G_4$ flux, determining the gaugings, can now vanish when restricted 
to $\cD$. 

We have also  given a Type IIA string description of this M5-brane 
instanton obtained at weak coupling. The M5-brane splits in this case into 
a system of D6-branes, D4-branes and an NS5-brane. 
The D6-branes and the NS5-brane are localized at points on the 4d/3d circle, while the D4-branes stretch between 
them along intervals. It was possible to 
connect the geometrical features of the gauge instanton bundle on Taub-NUT to 
wrapping numbers and relative positions of the Type IIA branes. It will be very interesting 
to extend this analysis of the M5-brane instanton and corresponding Type IIA configuration. 
In particular, the structure of zero modes have to be explored further to guarantee that 
the instantons actually correct the superpotential instead of other effective 3d couplings.

There are various further directions which can be explored to extend this 
work. As mentioned above, within the 4d/3d reduction one might hope
to get a better understanding of the backreacted 4d geometry of the effectively 
3d vacuum obtained by radion stabilization through Taub-NUT instantons. 
This also includes an extension to allow for a non-trivial warp factor both on the 4d/3d side,
as well as in the M-theory reduction. 
In particular, one might hope that the various 4d geometries parameterizing the points in  
the effective 3d potential might be found explicitly. 
The transition 
between the effectively 3d vacuum and the asymptotic 4d vacuum could 
admit an interpretation as phase transition in the 4d gravity theory 
as in \cite{Hawking:1982dh}. 

Independent of the precise interpretation 
one should be able to link the up-lifted 3d vacua with the 4d vacua 
by a transdimensional tunneling. In the stringy realization 
this corresponds to a tunneling from a 3d M-theory compactification 
to a 4d F-theory compactification. The 3d M-theory landscape will allow for 
more flux vacua then the 4d counterpart, and it deserves further study.
It would be also interesting to use the tunneling process to 
generate the initial conditions for an inflationary model realized in F-theory.   

On a more formal side, the analysis carried out in this paper reveals 
deep connections between consistency conditions of the string/M-theory 
compactifications and constraints on the fields found in the 
resulting 4d/3d supergravity theories \cite{GrimmTaylor}. In the study of consistent effective 
theories arising from string compactifications it will be important to 
also consider solutions of varying dimensionalities and topologies 
to constrain the couplings. A prominent role in our analysis 
took the couplings $\sigma,\tau$ of the $F^2$ and $\cR_4^2$ terms in the 4d effective theory as in \cite{GrimmTaylor}.
We have used circle fluxes to probe the connection of these couplings with 
the geometric data of the fourfold. It would be interesting to 
generalize this to include fluxes for the imaginary parts of 
$\sigma$ and $\tau$ since this will lead to genuinely 3d Chern-Simons gauge and gravity 
theories.

\section*{Acknowledgments}

We are thankful to Frederik Denef and Matthew Kleban for initial collaboration on this project. 
We gratefully acknowledge insightful discussions with Federico Bonetti, Frederik Denef, Mike Douglas, 
Eran Palti, Max Kerstan, Matthew Kleban, Wati Taylor, and Timo Weigand. TG acknowledges support and hospitality by 
the Simons Center, MIT, and the KITPC. RS would like to thank the Hong Kong Institute for Advanced 
Studies at HKUST for hospitality and financial support during part of this work.
This work was supported by a research grant of the Max Planck Society. 

\vspace*{1.5cm}

\appendix

\noindent {\bf \LARGE Appendices}

\section{4d/3d reduction with vectors}\label{Dualization4dVect}

In this appendix we add some more details about the 4d/3d reduction in the presence of 4d vectors, discussed in subsection \ref{4dVectors}.
A related analysis can be found with additional details in \cite{Grimm:2010ks,Grimm:2011tb}.

After integration of \eqref{4dkineticphiVect} over $S^1$ and performing the Weyl rescaling $g^{(3)}\rightarrow r^{-2}g^{(3)}=Rg^{(3)}$ in order to bring the 3d Einstein-Hilbert term to the canonical form, 
we find the 3d action in the vector multiplet formalism
\bea\label{3dkineticTVect}
S^{(3)}&=& \int-\cfrac12 R_3*_3{\bf{1}}+\cfrac14\tilde K_{L^IL^J}F^I\wedge *_3 F^J+\cfrac14\tilde K_{L^IL^J}dL^I\wedge *_3 dL^J \\ &&\quad +F^I\wedge{\rm Im}\,(\tilde K_{L^I\tau}d \tau) 
- \tilde K_{T_\alpha \bar T_{\bar\beta}} \nabla T_\alpha\wedge *_3  \overline{\nabla T_{\beta}} -(V^{3d}_{\rm F} + V^{3d}_{\rm D})*_3\bf{1}\,,
\eea
where $I=\{0,i\}$ and the kinetic potential $\tilde K$ is given in \eqref{KineticPotentialVect}.
Note that 
due to the non-trivial background fluxes \eqref{GaugeInvCircleFlux} the covariant derivatives in \eqref{gauging4D} are improved and become the three-dimensional covariant (or rather invariant) derivatives 
\beq\label{InvDerivatVect}
     \nabla T_\alpha = d T_\alpha +X_{i\alpha} A^i+ i  M_\alpha A^0\ .
\eeq
The D-term scalar potential obtained from the reduction of \eqref{4dkineticphiVect} is given by
\be\label{ScalPotRedVect}
V^{3d}_{\rm D}= 4K_{T_\alpha \bar T_{\bar\beta}}\left(R\Theta_{0\alpha}+\Theta_{i\alpha}\xi^i\right)\left(R\Theta_{0\beta}+\Theta_{j\beta}\xi^j\right)+2R({\rm Re}\,\tau)^{ij}\Theta_{i\alpha} \Theta_{j\beta}K_{T_\alpha}K_{T_\beta}\,.
\ee
It is easy to see that it coincides with the D-term potential in the 
general form \eqref{3dPotential}, with the $\cT$-potential given by 
eq. \eqref{ExplicitDtermPot} and the K\"ahler potential by eq. \eqref{KaehlerPotVect}. 
Indeed the first term in \eqref{ScalPotRedVect} arises from 
$K^{T_\alpha\bar T_{\bar\beta}}\partial_{T_\alpha}\cT\partial_{\bar T_{\bar\beta}}\cT$. 
As for the second term, one has
\be
K^{T_I\bar T_{\bar J}}\partial_{T_I}\cT\partial_{\bar T_{\bar J}}\cT=16\left(K_{T_I\bar T_{\bar J}}-K_{T_I}K_{\bar T_{\bar J}}\right)\Theta_{I\alpha}\Theta_{J\beta}K_{T_\alpha}K_{\bar T_{\bar\beta}}\,,
\ee 
and for the K\"ahler potential  \eqref{KaehlerPotVect} one finds 
\be
K_{T_I\bar T_{\bar J}}-K_{T_I}K_{\bar T_{\bar J}}=\cfrac18R({\rm Re}\,\tau)^{ij}\,.
\ee


\end{document}